\documentclass[usenatbib,onecolumn]{mnras}
\usepackage{graphicx}
\usepackage{amsmath,amssymb}
\usepackage{epsf}
\usepackage{dcolumn}
\usepackage{color}
\usepackage{epsf}
\usepackage{psfrag}
\input{epsf}

\newcommand{\beq}{\begin{equation}}
\newcommand{\eeq}{\end{equation}}

\newcommand{\ber}{\begin{eqnarray}}
\newcommand{\eer}{\end{eqnarray}}
\usepackage{graphicx}
\usepackage{adjustbox}
\usepackage{blindtext}

\def\beq{\begin{equation}}
\def\eeq{\end{equation}}

\def\ber{\begin{eqnarray}}
\def\eer{\end{eqnarray}}

\begin{document}

\title[Model independent constraints on dark energy]{Model independent constraints on dark energy evolution from low-redshift observations}

\author[Capozziello et al.]
{Salvatore Capozziello $^{1}$\thanks{E-mail:capozziello@na.infn.it}, Ruchika $^{2}$\thanks{E-mail:ruchika@ctp-jamia.res.in},
Anjan A Sen $^{2}$ \thanks{E-mail:aasen@jmi.ac.in} \\
$^{1}$Dipartimento di Fisica "E. Pancini", Universit\`a di Napoli  ``Federico II'', Via Cinthia, I-80126, Napoli, Italy.\\
$^{1}$Instituto Nazionale di Fisica Nucleare (INFN), Sez. di Napoli, Via Cinthia 9, I-80126 Napoli, Italy.\\
$^{1}$Gran Sasso Science Institute, Via F. Crispi 7, I-67100, L'Aquila, Italy.\\
$^{2}$Centre for Theoretical Physics, Jamia Millia Islamia, New Delhi-110025, India.}

%\pacs{04.50.-h, 04.20.Cv, 98.80.Jk}
%\keywords{Cosmological parameters,dark energy, cosmology:theory, cosmology:observations}

\maketitle
\date{today}

\begin{abstract}
Knowing the late time evolution of the Universe and finding out the  causes for this evolution are the important challenges of modern  cosmology. In this work, we adopt a model-independent cosmographic approach  and approximate the Hubble parameter considering the Pade approximation which works better than the standard Taylor series approximation for $z>1$. With this, we constrain the late time evolution of the Universe considering low-redshift observations coming from   SNIa, BAO, $H(z)$, $H_{0}$ , strong-lensing time-delay as well as the Megamaser observations for angular diameter distances. We confirm the tensions with $\Lambda$CDM model for low-redshifts observations. The present value of the equation of state for the dark energy has to be phantom-like and for other redshifts, it has to be either phantom or should have a phantom crossing. For lower values of $\Omega_{m0}$, multiple phantom crossings are expected. This poses serious challenges for single,  non-interacting scalar field models for dark energy. We derive constraints on the {\it statefinders} $(r,s)$ and these constraints show that a single dark energy model cannot fit  data for the whole redshift range $0\leq z\leq 2$: in other words,  we need multiple dark energy behaviors for different redshift ranges. Moreover, the constraint on sound speed for the total fluid of the Universe, and for the dark energy fluid (assuming them being barotropic), rules out the possibility of a barotropic fluid model for unified dark sector and barotropic fluid model for dark energy, as fluctuations in these fluids are unstable as $c_{s}^2 < 0$ due to constraints from low-redshift observations.
\end{abstract}

\section{Introduction}
The observed late time acceleration of the  Universe is one of the most important milestones for research in cosmology as well as gravitational physics. It behoves us to go beyond the standard attractive nature of gravity and compels us to think out of the box to explain the repulsive nature of gravity that is at work on large cosmological scales. Whether the reason for this repulsive gravity is due to the presence of non-standard component with negative pressure in the Universe ( {called \it dark energy}) \citep{ de1, de2, de3a, de3b, de4} or due to large scale (infrared) modification of Einstein's General Theory of Gravity (GR)  \citep{mod1, mod2, mod3, mod4, mod5, mod6, mod7, mod8, mod9}, is not settled yet. Still, recent results by Planck-2015 \citep{ ade1,ade2} for Cosmic Microwave Background Radiation (CMB), equally complimented by observations from Baryon-Acoustic-Oscillations (BAO) \citep{  bao1, bao2, bao3, bao4, bao5}, Supernova-Type-Ia (SNIa) \citep{ sn1, sn2, sn3}, Large Scale Galaxy Surveys (LSS) \citep{lss}, Weak-Lensing (WL) \citep{wl} etc, have put very accurate bound on the late time evolution of the Universe. It tells that the concordance $\Lambda$CDM Universe  is by far the best candidate to explain the present acceleration of the present Universe. But the theoretical puzzle continues to exists as we still do not know any physical process to generate a small cosmological constant, which is consistent with observations but is far too small (of the order of $10^{-120}$) compared to what one expects from standard theory of symmetry breaking. Problem like cosmic coincidence also remains. 

Interestingly,  few recent observational results have indicated inconsistencies in the cosmological credibility of the $\Lambda$CDM model. The model independent measurement of $H_{0}$ by Riess et al. (R16) \citep{R16} and its recent update \citep{R18}, have shown a tension of more than $3\sigma$ with the $H_{0}$ measurement by Planck-2015 for $\Lambda$CDM Universe (see also \citep{micol}). Similarly, mild inconsistency in $H_{0}$ for $\Lambda$CDM model is also observed by Strong Lensing experiments like H0LiCow using time delay \citep{holicow}. Subsequently \citep{valentino} have shown that such an inconsistency in $\Lambda$CDM model can be sorted out if one goes beyond cosmological constant and assumes dark energy evolution with time. In recent past, Sahni et al.\citep{sahni} have also confirmed this dark energy evolution in model independent way using the BAO measurement. More recently, using a combination of cosmological observations from CMB, BAO, SNIa, LSS and WL,  Zhao et al. \citep{gongbo} have shown (in a model independent way) that dark energy is not only evolving with time but it has also gone through multiple phantom crossings during its evolution. Such a result, if confirmed by other studies, is extremely interesting as it demands the model building scenario to go beyond single scalar field models which, by construction,  do not allow phantom crossing. It was first shown by Vikman \citep{vikman} and later by other authors in different contexts \citep{phcross1, phcross2,phcross3}. For a discussion on this topic in modified gravity, see \citep{bamba}.

Although the majority of the studies to constrain the late time acceleration of the Universe assumes a specific model (either for dark energy or for modification of gravity), there have been attempts to study such issues in a model independent way. But in most of these model independent studies, the goal is to reconstruct the dark energy equation of state \citep{emilio}. There are two major issues for such studies. Reconstructing the dark energy equation of states needs a very precise knowledge about the matter energy density in the Universe. A small departure from that precise value can result completely wrong result for dark energy equation of state\citep{om}. To avoid such issues, a number of diagnostics have been proposed that can independently probe dark energy dynamics without any prior knowledge about matter density or related  parameters. {\it Statefinders} \citep{state1, state2} and {\it Om Diagnostic}\citep{om} are some of the most interesting diagnostics. Moreover, in a recent study, the limitation of using the equation of state to parametrize dark energy has been discussed\citep{scherrer}. It is shown that  the energy density of dark energy or equivalently the Hubble parameter $H(z)$ are reliable observational quantities to distinguish  different dark energy models.

In the literature, there are different approaches for model independent study for late time acceleration in the Universe. Some of the most studied approaches are Principal Component Analysis (PCA) \citep{pc}, Generic Algorithm (GA) \citep{ ga1, ga2}, Gaussian Processes (GP)\citep{gp1,gp2} etc. See \citep{penn} for a nice review on different approaches for reconstruction. Cosmographic approach to constrain the background evolution of the Universe is a simple and yet useful approach\citep{cosmography}. Once the assumption of {\it cosmological principle} is made, it gives model independent limit for the background evolution of the Universe around present time. Given that the dark energy only dominates close to present time (except for early dark energy models), this approach is a powerful one to constrain the late time background evolution. The usual cosmographic approach involves Taylor expansion of cosmological quantities like scale factor and to define its various derivatives like Hubble Parameter ($H$), deceleration parameter ($q$), jerk ($j$), snap ($s$) and so on and subsequently constrain these parameters (at present time) using different cosmological  observations. This, in turn, can result constraints on the background evolution. This is completely model independent as no assumption of underlying dark energy model is needed. The major problem for this approach is that the Taylor expansion does not converge for higher redshifts and hence one cannot truncate the series at any order. To overcome this divergence problem for redshifts $z > 1$, the formalism of Pade Approximation (PA) for Cosmographic analysis was first proposed by Gruber and Luongo \citep{PA} and later by Wei et al. \citep{wei}. It was shown that the PA method is a better alternative than Taylor Series expansion, as the convergence radius of PA is larger than Taylor Series expansion. Hence, to constrain late time universe using Cosmography, PA is a reliable choice to extent the analysis to  higher redshifts.

In most of the analysis, PA is applied to write down the dark energy equation of state\citep{PA, wei, pade}. Although this is a reasonable approach to constrain the late time evolution of the Universe, using dark energy equation of state has its own problems as discussed in earlier paragraphs. PA has been also used to approximate the energy density for the dark energy \citep{ basilakos}. Recently, PA is used to approximate the luminosity distance $d_{L}$ which is a direct observable related to SNIa observations \citep{ salv1, salv2} . This gives a very clean constraints on late time Universe from SNIa observations as no further information is needed. But if one wants to use other observations related to background evolution, one first needs to calculate the Hubble parameter $H(z)$ from $d_{L}(z)$ and then use this $H(z)$ to construct other observables.

Given the fact that all the low redshift observables are constructed solely from $H(z)$, it is natural to use PA to approximate the $H(z)$ itself. Moreover, using PA for $H(z)$ directly, one can allow both dark energy and modified gravity to model the background expansion. 

In this work, we take into account this approach. We use $P_{22}$ (Pade Approximation of order (2,2)) for the Hubble parameter $H(z)$. Subsequently we use latest observational results from SNIa, BAO, Strong Lensing, H(z) measurements, $H_{0}$ measurements by HST as well as the angular diameter distance measurements by Megamaser Cosmology Project to constrain the late time evolution of the Universe. We first derive the observational constraints on various cosmographic parameters and later use those constraints to reconstruct the evolution of various equation of state, {\it statefinder diagnostics} and sound speed. 

In Sec. 2, we describe the cosmography and the Pade Approximation;  Sec. 3A is devoted to the  description of  different observational data used in the present study; in Secs. 3B-3E, we describe various results that we obtain in our study and finally in Sec. 4, we give a summary and outline perspectives of the method.

\section{Cosmography and Pade Approximation}

In Cosmographic terminology, the first five derivatives of scale factor $a(t)$  are defined as the Hubble parameter ($H$), the deceleration parameter ($q$), the jerk parameter ($j$), the snap ($s$)  and the lerk ($l$): 
\begin{equation}{}
H(t) = \frac{1}{a}\frac{da}{dt}           ;\\
\end{equation}{}
\begin{equation}
q(t) =-\frac{1}{a}\frac{d^{2}a}{dt^{2}}\left[\frac{1}{a}\frac{da}{dt}\right]^{-2};
j(t) = \frac{1}{a}\frac{d^{3}a}{dt^{3}}\left[\frac{1}{a}\frac{da}{dt}\right]^{-3}
\end{equation}{}
\begin{equation}
s(t) = \frac{1}{a}\frac{d^{4}a}{dt^{4}}\left[\frac{1}{a}\frac{da}{dt}\right]^{-4};
l(t) = \frac{1}{a}\frac{d^{5}a}{dt^{5}}\left[\frac{1}{a}\frac{da}{dt}\right]^{-5}
\end{equation}{}

\noindent
The Taylor Series expansion of the Hubble parameter around present time ($z=0$) is:
\begin{equation}{}
H(z) = H_{0}+H_{10} z+ \frac{H_{20}}{2!} z^{2}+...           ;\\
\end{equation}
 where $H_{i0} = \frac{d^{i}H}{dz^{i}}|_{z=0}$. Here the derivatives of Hubble parameter can be expressed as
\begin{equation}\nonumber
H_{1}=H_{10}/H_{0}= 1+q_{0}
\end{equation}
\begin{equation}\nonumber
H_{2}=H_{20}/H_{0}= -q_{0}^{2}+j_{0}
\end{equation}
\begin{equation}\nonumber
H_{3}=H_{30}/H_{0}= 3q_{0}^{2}(1+q_{0})-j_{0}(3+4q_{0})-s_{0}
\end{equation}
\begin{equation}
H_{4}=H_{40}/H_{0}= -3q_{0}^{2}(4+8q_{0}+5q_{0}^{2}) +j_{0}(12+32q_{0}+25q_{0}^{2}-4j_{0})+ s_{0}(8+7q_{0})+l_{0}
\end{equation}

The series (4) does not converge for redshift $|z|>$1. So, to increase the radius of convergence, we use Pade Approach (PA). The PA is developed using the standard Taylor Series definition but it allows better convergence at higher redshifts.We define the (N,M) order PA as the ratio:

\begin{equation}
P_{NM} = \frac{\displaystyle\sum_{n=0}^{N} a_{n}z^{n}} {1+\displaystyle \sum_{m=1}^{M} b_{m}z^{m}}.
\end{equation}

\noindent
$P_{NM}$ has total $(N+M+1)$ number of independent coefficients. One can Taylor Expand $P_{NM}$ and equate the coefficients to that for a generic function expanded as power series $f(z) =\displaystyle \sum_{i=0}^{\infty} c_{i} z^{i}$ to get
\begin{equation}\nonumber
P_{NM}(0) = f(0)
\end{equation}
\begin{equation}\nonumber
P'_{NM}(0) = f'(0)
\end{equation}
\begin{equation}
P''_{NM}(0) = f''(0)
\end{equation}
\begin{equation}
.......
\end{equation}
\begin{equation}\nonumber
P^{N+M}(0) = f^{N+M}(0).
\end{equation}

Hence one can always write any function expanded in Taylor Series in terms of PA as:

\begin{equation}
f(z) = \displaystyle\sum_{i=0}^{\infty} c_{i} z^{i} =  \frac{\displaystyle\sum_{n=0}^{N} a_{n}z^{n}} {1+ \displaystyle\sum_{m=1}^{M} b_{m}z^{m}} + O(z^{N+M+1}).
\end{equation}

\noindent
In this work, we use $P_{22}$ to approximate the Hubble parameter $H(z)$. As mentioned in the Introduction, all the observables related to low-redshift observations are directly related to $H(z)$ or they directly measure $H(z)$. So it is more reasonable that we use PA for $H(z)$ itself. So we assume:

\begin{equation}
E(z) = \frac{H(z)}{H_{0}} = \frac{P_{0}+P_{1}z+P_{2}z^{2}}{1+Q_{1}z+Q_{2}z^{2}}.
\end{equation}
 
\noindent
Remember that the ``normalised Hubble Parameter $E(z)$" is present in different expressions for observables like luminosity distance $d_{L}$, angular diameter distance $d_{A}$ etc. Hence we apply PA to $E(z)$. While choosing Pade orders, we keep in mind that
\begin{itemize}
\item The Pade function[NM] should smoothly evolve in all redshift ranges used for cosmographic analysis.
\item All Pade Approximations used, should give Hubble parameter positive.
\item The degree of polynomials for numerator and denominator should be close to each other.
\item While using a combination of datasets, the cosmographic priors should be chosen so that it does not provide divergences.  
\end{itemize}

\begin{figure}
\begin{center}
\resizebox{250pt}{200pt}{\includegraphics{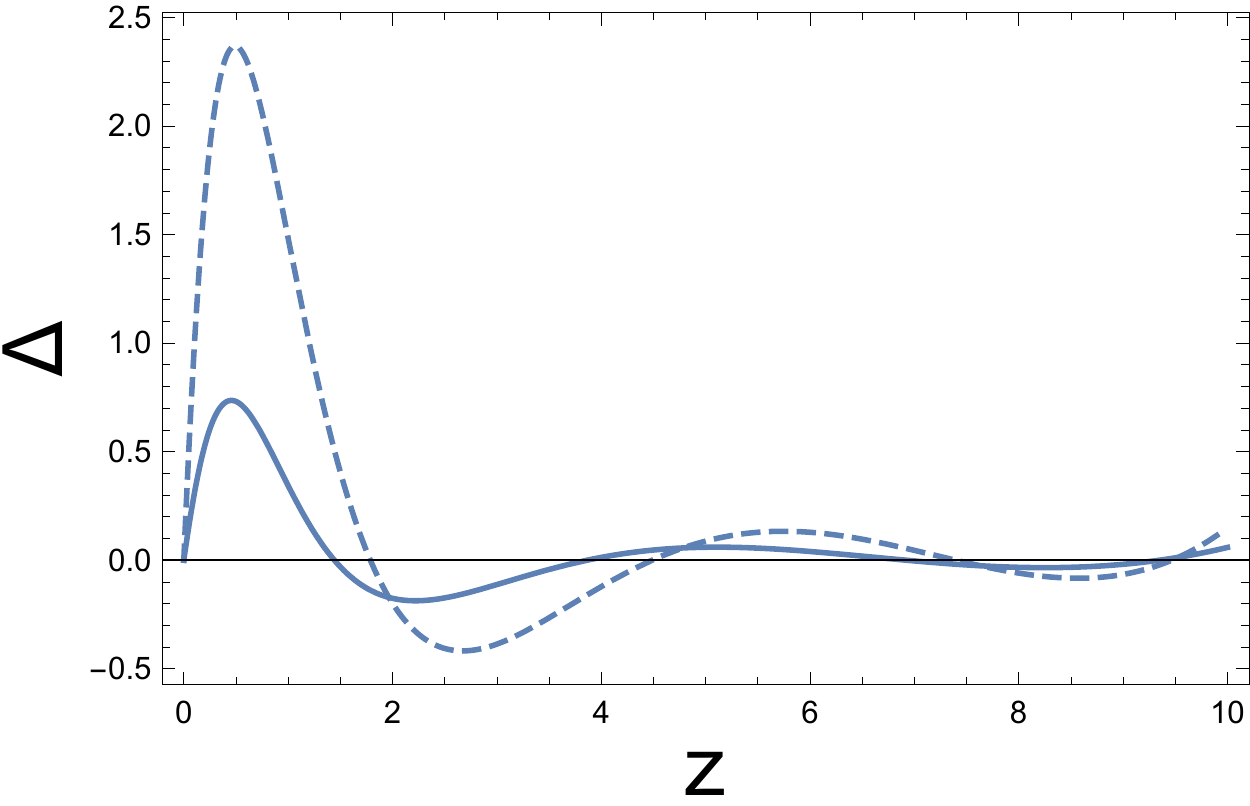}} 
\end{center}
\caption{ The percentage difference ($\Delta$) between the actual model (see text) and fitted model as a function of redshift. Dashed line is fourth order Taylor Series Expansion and solid line is for $P_{22}$.}
\end{figure}

One can always redefine the parameters in (10) and set $P_{0} =1$ so that $H(z=0) = H_{0}$. Furthermore the parameters $P_{1}, P_{2}, Q_{1}$ and $Q_{2}$ are not physically relevant parameters. One can always relate them to physically relevant kinematic quantities like $q_{0}, j_{0}, s_{0}, l_{0}$ which represent cosmographic quantities for the cosmological expansion. For this, one needs to take different derivatives for $H(z)$ given in (10) at $z=0$ and relate them to  $q_{0}, j_{0}, s_{0}, l_{0}$ and  solve $P_{1}, P_{2}, Q_{1}$ and $Q_{2}$ in terms of $q_{0}, j_{0}, s_{0}, l_{0}$. Subsequently one gets:

\begin{equation}\nonumber
P_1 =  H_1 + Q_1,
\end{equation}
\begin{equation}\nonumber
P_2 = \frac{H_2}{2} + Q_1H_1 +Q_2,
\end{equation}
\begin{equation}\nonumber
Q_1 = \frac{-6 H_1 H_4 +12 H_2 H_3}{24 H_1 H_3 - 36 H_2^2},
\end{equation}
\begin{equation}
Q_2 = \frac{3 H_2 H_4 -4 H_3^2}{24 H_1 H_3 - 36 H_2^2},
\end{equation}

\noindent
where $H_{1}, H_{2}, H_{3}$ and $H_{4}$ are related to cosmographic pramaters ($q_{0}, j_{0}, s_{0}, l_{0}$) in equation (5).  

Before studying the observational constraints, we compare (10) with a fourth order Taylor Series expansion for $E(z)$ which also contains four arbitrary parameters. We fit both to a $\Lambda$CDM model given by 
$E(z) = \sqrt{0.3 (1+z)^3 + 0.7}$. In Figure 1, we show which one fits the actual model better by plotting the difference between the fit and the actual model. One can clearly see, that in the redshift range $0 \leq z \leq 2$, where the effects of repulsive gravity  (and hence the late time acceleration) is dominant, the PA gives much better fit to the actual model than the Taylor series expansion and for PA, the deviation from actual model is always less than a percent. One should note that most of the low-redshift observational data are in this redshift range. 

Furthermore, as one can see from Figure 1, the deviation from the actual model (in this case, from $\Lambda$CDM) is maximum around $z\sim 0.5$, and this is true for Taylor series expansion as well as Pade. Given the fact that a large number of observational data are present around this redshift, it raises the obvious question that whether Pade is a good parametrisation to represent actual model around this redshift. But as one can see, the maximum deviation from the actual model, in case of Pade, is always less than $1\%$ although it is around $2.5\%$ for Taylor series expansion. As the accuracy of present observational data related to background universe, is still greater than $1\%$, we should not worry about this peak in deviation $\Delta$ around $z\sim 0.5$.  We show in subsequent sections that Pade can indeed put sufficiently strong constrains on parameters related to background expansion around $z\sim 0 - 2$, which is enough to rule out a large class of standard dark energy behaviours.

\section{Observational Constraints}
\subsection{Observational Data}

To obtain observational constraints for six arbitrary parameters (four parameters in the expression for $E(z)$ together with present day Hubble parameter $H_{0}$ and the sound horizon at drag epoch $r_{d}$ related to BAO measurements  ) we use the following low-redshifts datasets involving background cosmology:

\begin{itemize}
\item The isotropic BAO measurements  from 6dF survey, SDSS data release for main galaxy sample (MGS) and eBoss quasar clustering as well as from Lyman-$\alpha$ forest samples. For all these measurements and the corresponding covariance matrix, we refer readers to the recent work by Evslin et al. \citep{evslin} and references therein.

\item Angular diameter distances measured using water megamasers under the Megamaser Cosmology Project \citep{evslin, maser1, maser2, maser3}.

\item Strong lensing time-delay measurements by H0LiCOW experiment (TDSL) \citep{holicow}.

\item The OHD data for Hubble parameter as a function of redshift as compiled in Pinho et al  \citep{ ohd}.

\item The latest measurement of $H_{0}$ by Riess et al (R16) \citep{R16}.

\item Latest Pantheon data for SNIa in terms of $E(z)$  \citep{pantheon1, pantheon2}.

\end{itemize}

\subsection{Results}

To carry out the detail statistical analysis to find the constraints on the cosmological expansion, we can proceed in two ways. We can directly use equation (10) with $ P_{0} =1$ and find constraints on $H_{0}, P_{1}, P_{2}, Q_{1}, Q_{2}$, or we can use the relations between  $P_{1}, P_{2}, Q_{1}, Q_{2}$ and $q_{0}, j_{0}, s_{0}, l_{0}$ as given in (11) and use $q_{0}, j_{0}, s_{0}, l_{0}$ as parameters in our model. We study both the cases and find constraints on $H(z)$. The reconstructed $H(z)$ in these two cases is shown in Figure 2.  As one can see, both approaches have similar results for redshifts $z<1$. But for higher redshifts (in particular for $z = 1.5$ and above), there are disjoint regions between two cases beyond $68\%$ confidence interval, showing some amount of disagreement between the two approaches.

In our following calculations, we adopt the second approach where we use the parameters $q_{0}, j_{0}, s_{0}, l_{0}$ for our study as these are directly related to the cosmological expansion. As an example, assuming  $j =1$ at all redshifts, directly implies the $\Lambda$CDM behaviour. Similarly the sign change  of $q$ parameter defines the deceleration-to-acceleration transition. There is no direct physical interpretation for the set of parameters  $P_{1}, P_{2}, Q_{1}, Q_{2}$.

\begin{figure}
\begin{center}
\resizebox{250pt}{200pt}{\includegraphics{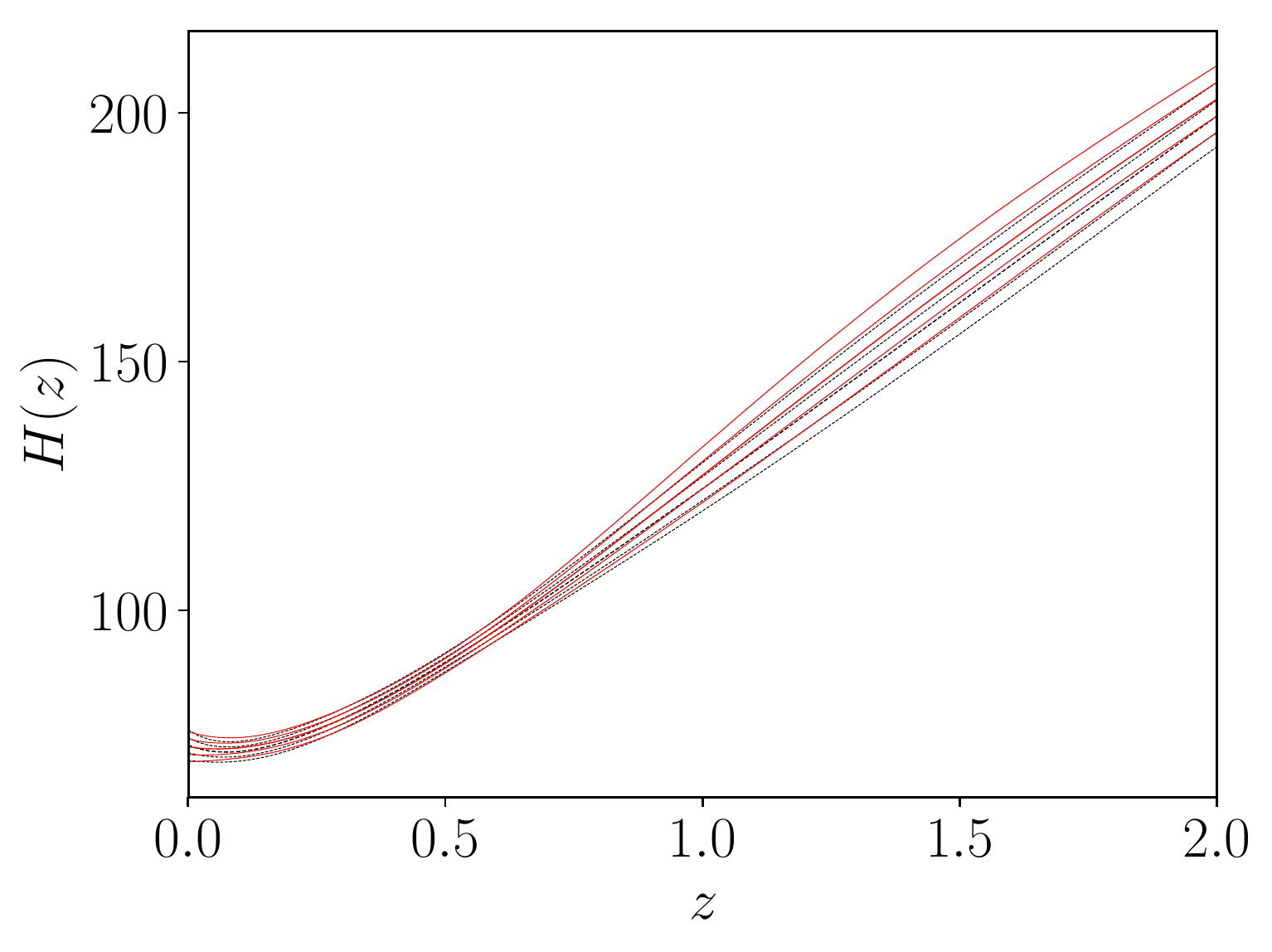}} 
\end{center}
\caption{ Reconstructed $H(z)$ using observational data (see text). The solid lines are for using $q_{0}, j_{0}, s_{0}, l_{0}$ as parameters and the dashed lines are for using  $P_{1}, P_{2}, Q_{1}, Q_{2}$ as parameters. In both cases, the innermost line is for the best fit case and the other two sets are for $68\%$ and $95\%$ confidence level.}
\end{figure}

The best fit and $1\sigma$ bounds on the parameters $s_{0}$ and $l_{0}$ for combination of all the datasets are $19.97^{+11.57}_{-10.84}$ and $121.41^{+91.94}_{-83.56}$ respectively showing that we do not have strong constraints for these two parameters from the current datasets.

\begin{figure}
\begin{center}
\resizebox{380pt}{350pt}{\includegraphics{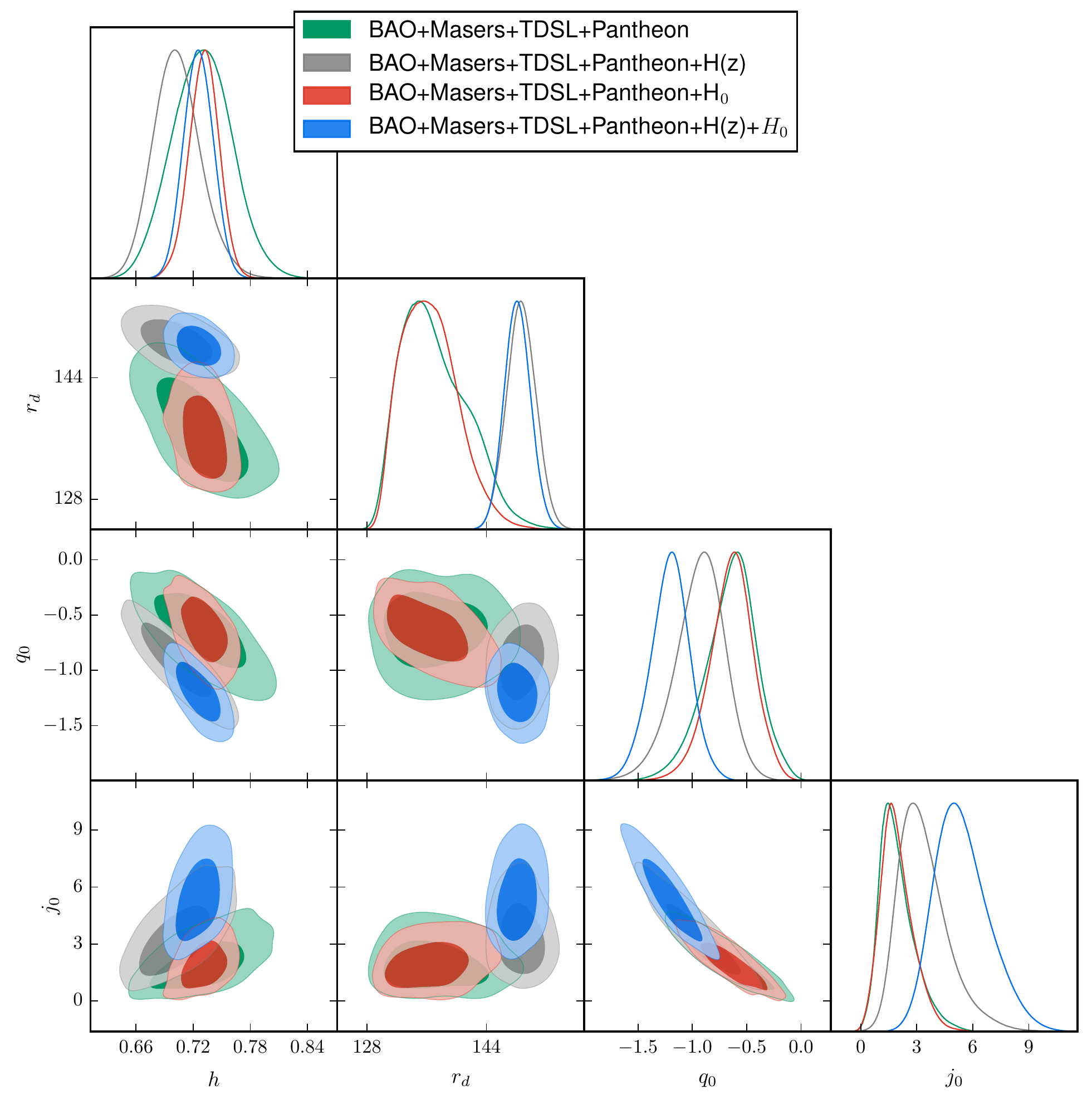}} 
\end{center}
\caption{The likelihoods for different cosmographic parameters and $r_{d}$ as well as the confidence contours in different parameter space. For each contour, the deep shaded region is for $68\%$ confidence level and light shaded region is for $95\%$ confidence level.}
\end{figure}

In Figure 3, we show the likelihoods and confidence contours for the rest of the cosmographic parameters, e.g., $h, q_{0},  j_{0}$ and the sound horizon at drag epoch $r_{d}$ that appears in the BAO measurements. Please note that in our calculations, we assume $H_{0} = 100 h$  Km/s/Mpc. The best fit values along with $1\sigma$ bounds for these parameters are summarised in Table I. As one can see, although we can not put strong bounds on parameters $s_{0}$ and $l_{0}$ ( which are related to fourth and fifth order derivative of the scale factor), but the constraints on $h$, $q_{0}$ and $j_{0}$ are pretty tight. In other words, the Pade can tightly constrain upto third derivative of the scale factor. This, in turn, can give very strong constraints on other derived parameters like $w_{eff}$,  $w_{DE}$ as well as different statefinder parameters, as shown in subsequent sections. 

%\begin{table}
%% use tabular font for a smaller size font
%\caption{Maximum Likelihood values and 1D marginalised $68\%$ confidence intervals of parameters for respective datasets.}\label{tab:bestfit}
\begin{table*}
\caption{ Maximum Likelihood values and 1D marginalised $68\%$ confidence intervals of parameters for respective datasets.}
%\centering
\begin{adjustbox}{max width=0.85\textwidth}
\begin{tabular}{|c|c|c|c|c|}
\hline
 & $ BAO+Masers+TDSL+Pantheon$ & $BAO+Masers+TDSL+Pantheon+H_{0}$ & $BAO+Masers+TDSL+Pantheon+H(z)$   & $BAO+Masers+TDSL+Pantheon+H_{0}+H(z)$\\
\hline
$h$  & $0.7293\pm 0.031$ &$ 0.7313 \pm 0.015$ & $0.7034\pm 0.024$ &$ 0.7256 \pm 0.015$ \\
$r_d$          & $137.06\pm 4.58$ & $136.41\pm 3.82$   & $148.67\pm 1.93$   &$ 148.16 \pm 1.74$ \\
$q_{0}$  & $-0.644 \pm 0.223$ & $-0.6401 \pm 0.187$  & $-0.930 \pm 0.218$ &$ -1.2037\pm 0.175$ \\
$j_{0}$     & $1.961^{+0.926}_{-0.884}$  & $1.9461^{+0.871}_{-.816}$ & $3.369^{+1.270}_{-1.294} $  &$5.423^{+1.497}_{-1.443}$ \\
\hline
\end{tabular} 
%\caption{Maximum Likelihood values and 1D marginalised $68\%$ confidence intervals of parameters for respective datasets.}
%\label{tab:bestfit} 
\end{adjustbox}
\end{table*}
%\end{table}

Following are the main results from Table I and Figure 3:

\begin{itemize}

\item  The low redshift measurements from BAO, Strong lensing, SNIa and angular diameter distance measurement by megamaser project,  give constraint on $H_{0}$ which is fully consistent with R16 constraint on $H_{0}$. Hence, in a model independent way, using low redshift observations, we confirm the consistency with $H_{0}$ measurement by R16 \citep{R16}. If one adds $H_{0}$ measurement by R16 to this combination, the best fit value for $H_{0}$ shifts to the higher side resulting tensions with Planck-2015 measurements  \citep{ ade1} for $\Lambda$CDM at $3.8\sigma$; in contrast, adding  $H(z)$ measurements to this combination, the best fit value for $H_{0}$ shifts to lower side and  tension with Planck-2015 measurement reduces to less than $2\sigma$. With combinations of all the data, the tension in $H_{0}$ with Planck-2015 result for $\Lambda$CDM is $3.46\sigma$. This result is completely model independent.

\item Without $H(z)$ data, the allowed value for sound horizon at drag epoch $r_{d}$ as constrained by low-redshift data, is substantially smaller than $r_{d}$ from Planck-2015 for $\Lambda$CDM model  \citep{ ade1}. This is consistent with the recent result obtained by Evslin et al.\citep{evslin} using different dark energy models. Here we obtain the same result in a model independent way. But adding $H(z)$ data shifts the constraint on $r_{d}$ on higher values making it consistent with the Planck-2015 measurement for $\Lambda$CDM. With combination of all the datasets, the measured value of $r_{d}$ is fully consistent with Planck-2015 results for $\Lambda$CDM. Consistency with Planck-2015 result for $r_{d}$ crucially depends on inclusion of $H(z)$ measurements. Without $H(z)$, there is nearly $3\sigma$ inconsistency ($2.9\sigma$ to be precise) with our model independent measurement for $r_{d}$ and that by Planck-2015.

\item With only low redshift measurements (BAO+SNIa+TDSL+Masers+$H(z)$+$H_{0}$), in a model independent way, we put strong constraint on deceleration parameter $q_{0}$  and it unambiguously confirms the late time acceleration.

\item One interesting result is for the jerk parameter $j$. For $\Lambda$CDM, $j = 1$ always. Our result shows that  $j_{0} =1$ is ruled out at  $3.06\sigma$ confidence limit.  There is a similar tension between  $H_{0}$ measurement by R16 and Planck-2015 constraint on $H_{0}$ for $\Lambda$CDM \citep{ R16, R18}. Our study confirms similar tension with low redshifts measurements in a model independent way in terms of the jerk parameter $j_{0}$.

\item From the likelihood plots and confidence contours in Figure 3, it is interesting to see that $H(z)$ data pulls the results away from what one gets with rest of the datasets. This is true for all the parameters. The result for combination of all datasets strongly depends on $H(z)$ data. We stress that this result is independent of any underlying dark energy or modified gravity models.

\end{itemize}

\subsection{The role of the Equations of State}

In the absence of spatial curvature, total energy density $\rho_{T}$ and the total pressure ($P_{T}$) of the  background Universe can be expressed as:

\begin{equation}
\rho_{T} = \frac{3 H^2 (z)}{8 \pi G} ; \quad P_{T} = \frac{H^2}{4 \pi G} \left(q(z)- \frac{1}{2}\right)
\end{equation}

\noindent
where $q(z)$ is the deceleration parameter of the Universe at any redshift. Using these, the effective equation of state of the background Universe is given by 
\begin{equation}
w_{eff}(z)= \frac{P_{T}}{\rho_{T}}  = \frac{2q(z)-1}{3}.
\end{equation}

Using the constraints on the cosmographic parameters, we reconstruct the behaviour $w_{eff}(z)$ and its $1\sigma$ and $2\sigma$ allowed behaviors are shown in Figure 4. To compare with the Planck-2015 results for $\Lambda$CDM\citep{ade1}, we also show the reconstructed $w_{eff}(z)$ for $\Lambda$CDM model using the constraints from Planck-2015. The figure shows the clear tension between the model independent $w_{eff}(z)$ behavior as constrained by low redshift data and the $w_{eff}(z)$ for $\Lambda$CDM model as constrained by Planck-2015. 

Note that, for an accelerating Universe, we should have $w_{eff} < -1/3$. One can see in Figure 4, as one goes to past, the universe exits the accelerating regime and enters the decelerated period at around $z \sim 0.6$. But it also allows another accelerating period at high redshifts ($z \sim 1.5$ and higher) although decelerated universe is always allowed for $z > 0.6$. Adding data from high redshifts observations like CMB may change this behavior at high redshifts. 

\begin{figure}
\begin{center}
\resizebox{260pt}{220pt}{\includegraphics{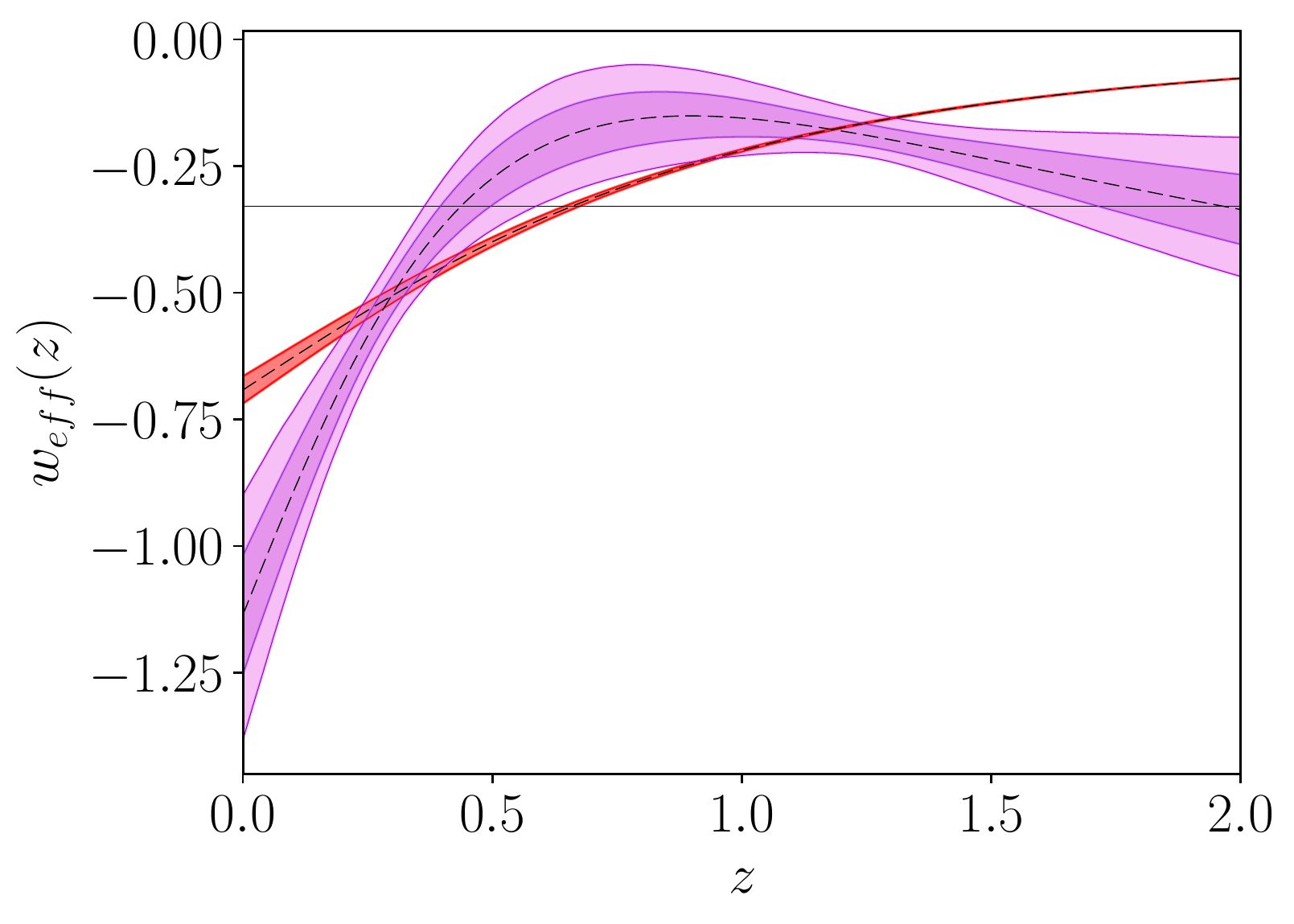}} 
\end{center}
\caption{Reconstructed  $w_{eff}$ as a function of redshift $z$. Pink shaded region is for model independent study in this paper. The red shaded region is for $\Lambda$CDM model as constrained by Planck-2015. The horizontal line is for $w_{eff} = -1/3$. The deep shaded and light shaded regions are for $68\%$ and $95\%$ confidence level. }\label{fig:rs_z}
\end{figure}

Although using (10), we constrain the overall background evolution of the Universe and it does not depend on any particular dark energy or modified gravity model, one can use the constraint on $E(z)$ to reconstruct the dark energy equation of state $w_{DE}(z)$.  Assuming that the late time Universe contains the non-relativistic matter and dark energy, the dark energy equation of state $w_{DE}(z)$ can be written as:

\begin{equation}
w_{DE}(z)= \frac{E(z)^{2}(2q(z)-1)}{3(E(z)^{2} - \Omega_{m0}(1+z)^3)}.
\end{equation}

\begin{figure}
\begin{center} 
\resizebox{220pt}{180pt}{\includegraphics{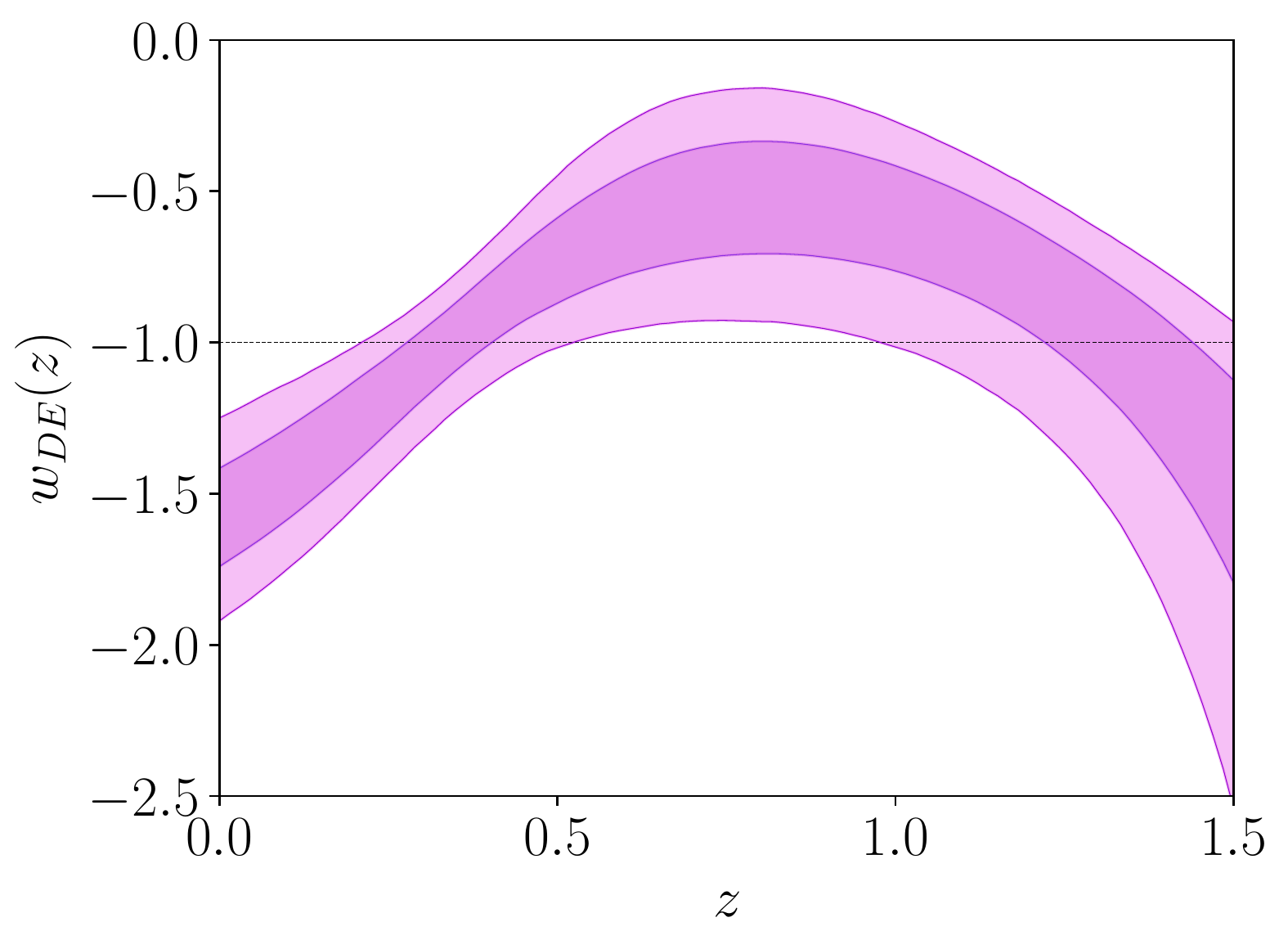}} 
\hspace{1mm} \resizebox{220pt}{180pt}{\includegraphics{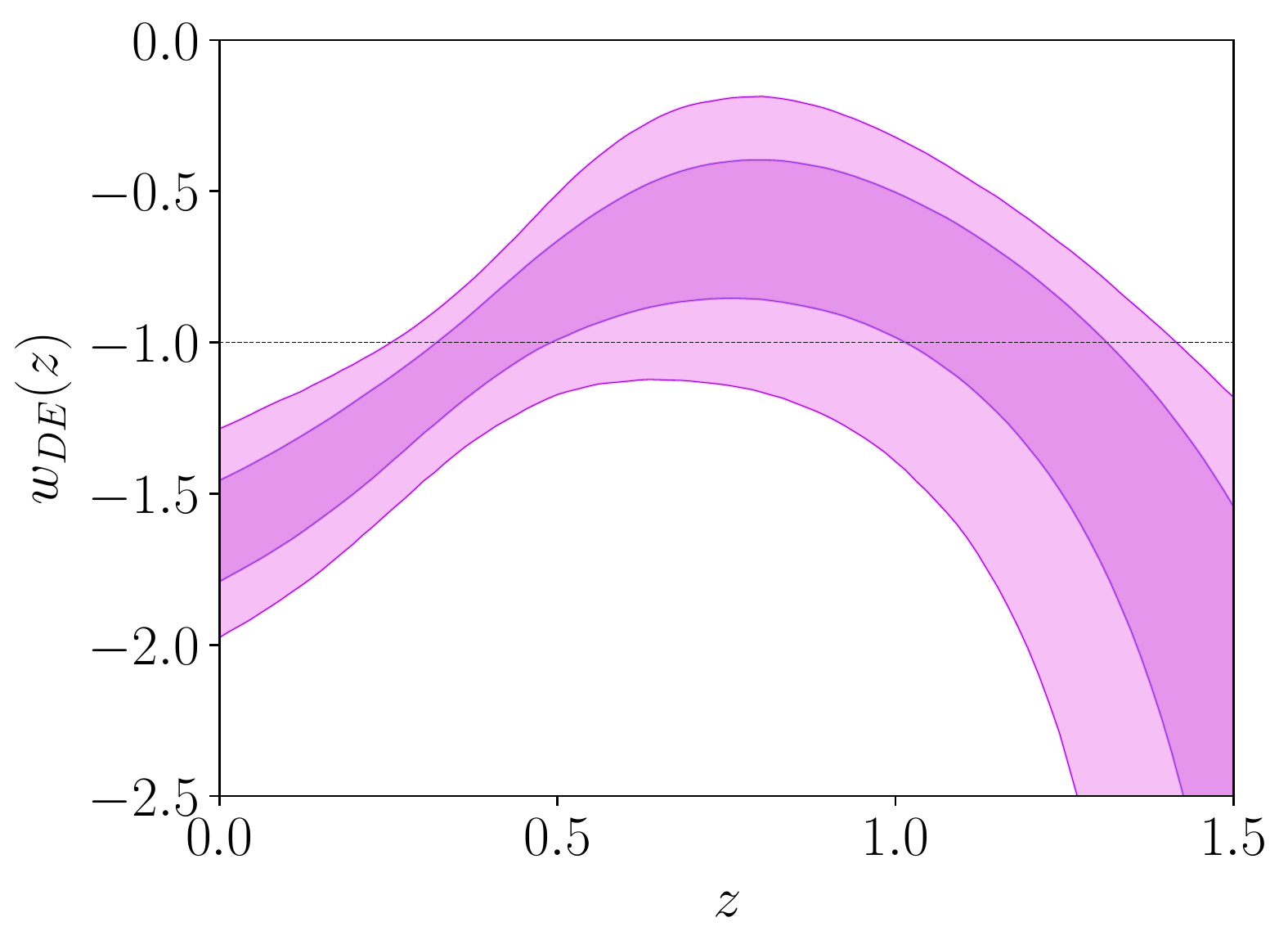}} 
\vspace{1mm} \resizebox{220pt}{180pt}{\includegraphics{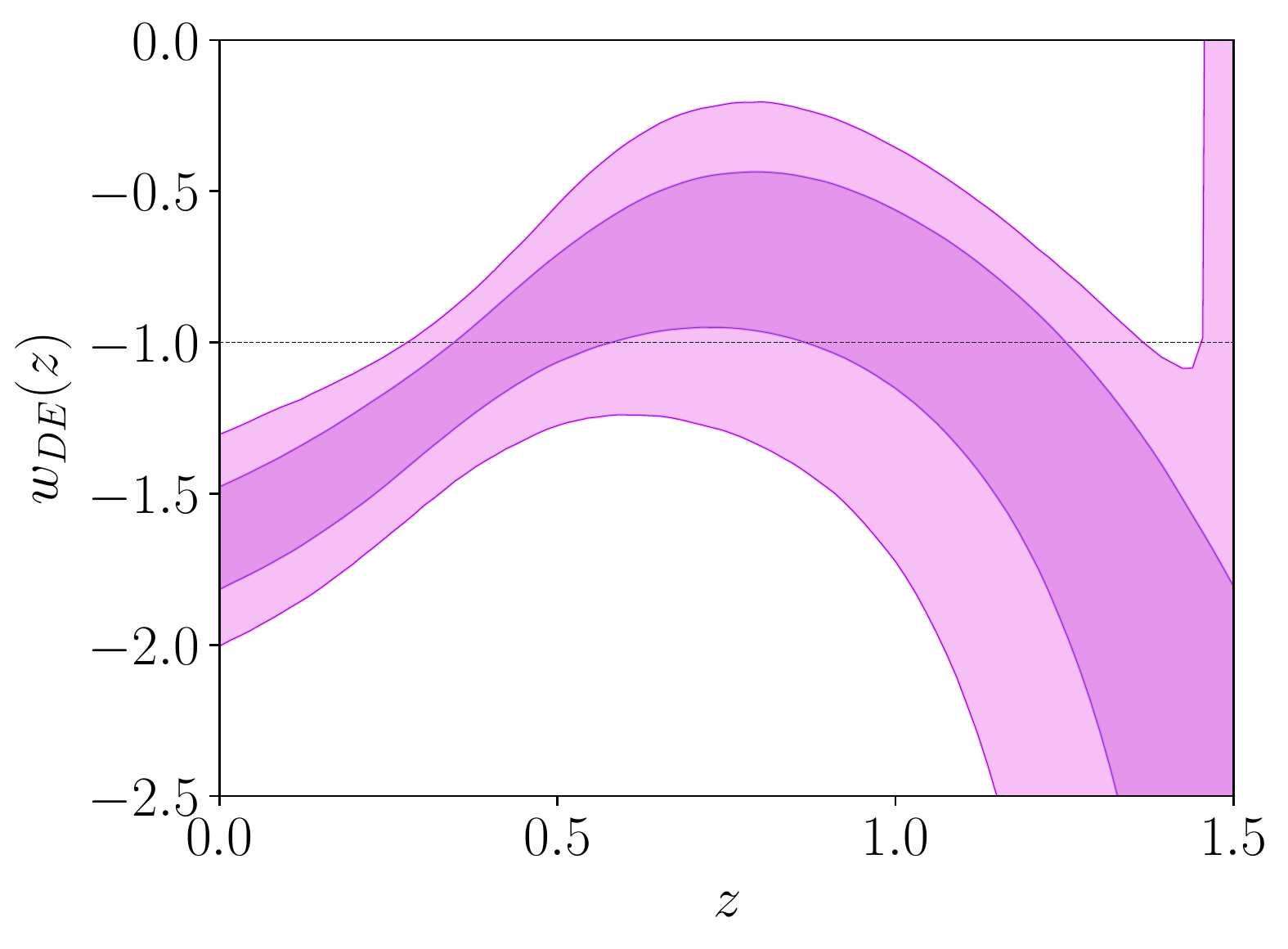}}
\end{center}
\caption{Reconstructed equation of state for dark energy $w_{DE}(z)$ taking values of $\Omega_{m0}$ =0.28(upper left), 0.30(upper right), 0.31(lower) respectively. Contour shadings are the same as in Figure 4.}\label{fig:wplot}
\end{figure}
 
\noindent
Here $\Omega_{m0}$ is the present day energy density parameter associated with the non-relativistic matter. As evident from this equation, one needs the information about $\Omega_{m0}$ to reconstruct $w_{DE}(z)$. In our model independent approach using PA for $E(z)$, $\Omega_{m0}$ is not a parameter that we fit with observational data. So in order to reconstruct $w_{DE}(z)$ using (14), we assume three values for $\Omega_{m0}$, e.g ($0.28,~0.30,~0.31$). The model independent reconstructed $w_{DE}(z)$ is shown in Figure 5. One can observe following results from these plots:

\begin{itemize}

\item For all three values of $\Omega_{m0}$, cosmological constant $(w_{DE} =-1$) is in more than $2\sigma$ tension with the reconstructed $w_{DE}(z)$ around present day. 

\item For $\Omega_{m0}=0.28$, this tension is also present for redshift range $0.4 \leq z \leq 1$.

\item For $\Omega_{m0} = 0.3$ and $0.31$, around $z\sim 1.5$, there is also $2\sigma$ inconsistency with $w=-1$.

\item Irrespective of the value for $\Omega_{m0}$, at low redshifts, the data allow only phantom behaviour ($w < -1$) for $w_{DE}$.

\item For $\Omega_{m0} = 0.28$, a pure phantom or pure non-phantom $w_{DE}$ is not allowed at all redshifts. There should be phantom crossing ( probably more than one).

\item For higher values of $\Omega_{m0}$, although pure phantom behaviour is  allowed at all redshifts, pure non-phantom behaviour is still not allowed at all redshifts. 

\item Keeping in mind that a single canonical and minimally coupled scalar field model only results non phantom dynamics, our results show that all such scalar field models are in tension with low redshift observations. And this result is independent of choice of any dark energy model.

\item The overall behaviour of reconstructed $w_{DE}(z)$ for any values of $\Omega_{m0}$, shows that the data prefer phantom crossing at low redshifts. As we discuss in the Introduction, crossing phantom divide with a single fluid non-interacting scalar field is difficult to achieve \citep{ vikman, phcross1, phcross2, phcross3} ; our results show that such models are in tension with low-redshift observations. However, in case of an imperfect non-canonical scalar field model, this problem may be avoided,  as shown by \citep{deff}.

\end{itemize}

\subsection{The Statefinder Diagnostics}

Till now, we have constrained the cosmographic parameters from low-redshifts data in a model independent way. This, in turn, allows us to reconstruct the total equation of state $w_{eff}$ of the Universe as well as the dark energy equation of state $w_{DE}$ (assuming that late time acceleration is caused by dark energy) for specific choices of $\Omega_{m0}$. But this does not allow us to pin point the actual dark energy model as there is a huge degeneracy between cosmographic parameters and different dark energy models. One needs some further model independent geometrical quantities that shed light on actual model dependence for dark energy. \citep{state1} have introduced one such sensitive diagnostic pair $(r,s)$, called ``{\it Statefinder Diagnostics}"  \citep{ state1, state2}.  They are defined as:

\begin{equation} 
r = \frac{\dddot a}{aH^3} = q^2 + \frac{H^{''}}{H} (1+z)^2,
\end{equation}
\begin{equation} 
s = \frac{r-1}{3(q-0.5)} .
\end{equation}

\noindent
Here ``prime" represents the derivative with respect to redshift. Remember the {\it statefinder} $r$ is same as the {\it jerk parameter} $j$ defined earlier. But we keep the original notation of {\it statefinders} as proposed by Sahni et al. \citep{state1}. One of the main goal of constructing any diagnostic is to distinguish any dark energy model from $\Lambda$CDM and {\it statefinders} $(r,s)$ do exactly this as for $\Lambda$CDM $(r,s) = (1,0)$ for all redshifts. Any deviation from this fix point in $(r,s)$ plane, signals departure from $\Lambda$CDM behaviour. Moreover the different trajectories in $(r,s)$ plane indicate different dark energy models including scalar field models with different potentials, different parametrizations for dark energy equation of state and even brane-world models for dark energy ( we refer readers to Figure 1 and Figure 2  \citep{state2}).

%%%
%%\begin{figure}[H]
%%\begin{center} 
%%\resizebox{220pt}{180pt}{\includegraphics{r-s.pdf}} 
%%\hspace{1mm} \resizebox{220pt}{180pt}{\includegraphics{cont_s_q.pdf}} 
%%\vspace{1mm} \resizebox{220pt}{180pt}{\includegraphics{cont_r_q.pdf}}
%%\end{center}
%%\caption{Upper left plot shows how r-s contour shift by varying redshift, Upper right plot shows the 3d plot of present value of cosmological parameters $q_0$, $j_0$ and $r_d$.Bottom plot is 3d plot showing how present hubble parameter changes with varing $q_0$ and $r_d$}\label{fig:3d_plot}
%%\end{figure}
%%%

We reconstruct the behaviours of $(r,s)$ and show different aspects of these reconstruction in Figure 6. Following are the results that one can infer from these plots:

\begin{figure}
\begin{center}
\resizebox{210pt}{180pt}{\includegraphics{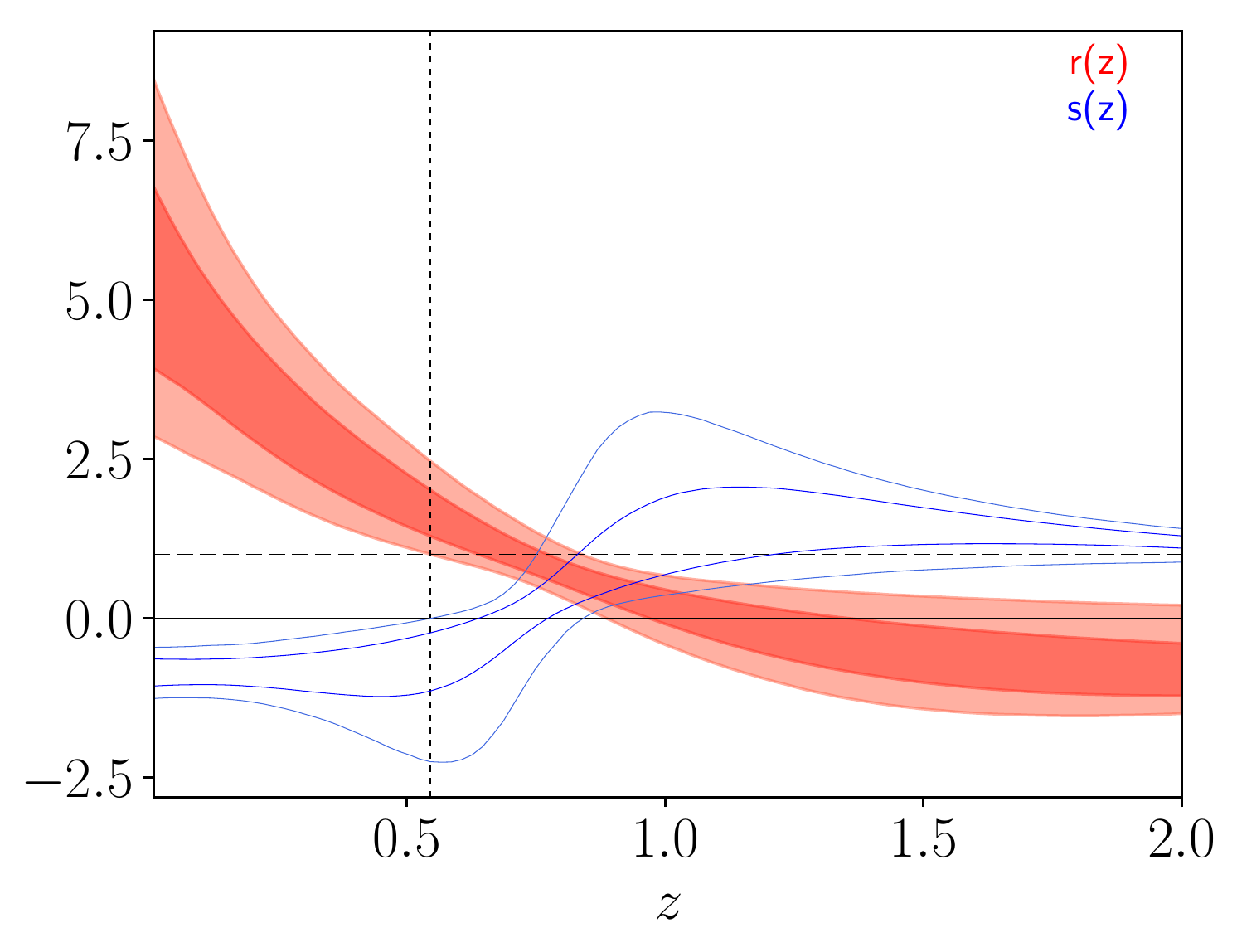}}
\hspace{1mm}\resizebox{220pt}{180pt}{\includegraphics{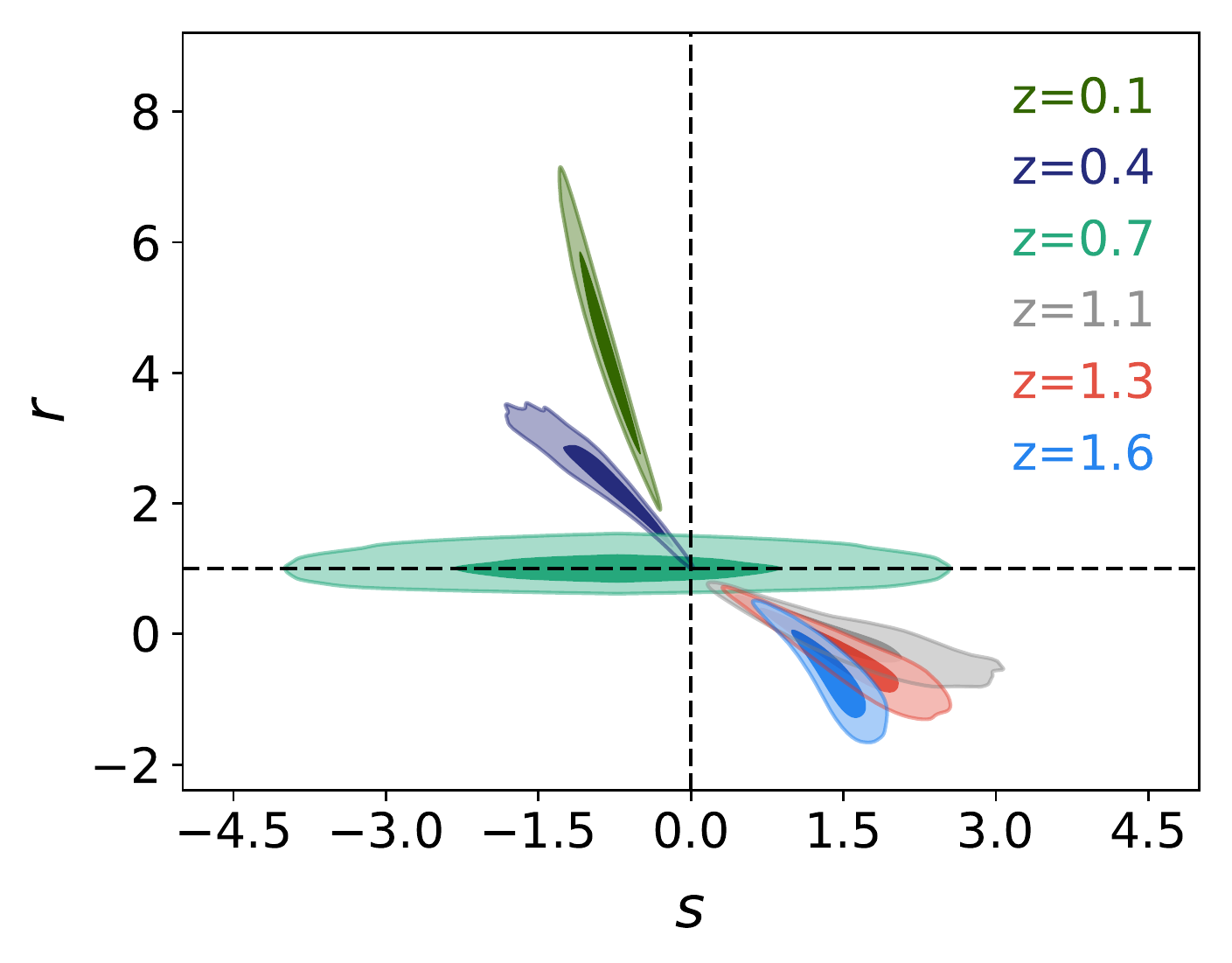}}\\
\resizebox{220pt}{180pt}{\includegraphics{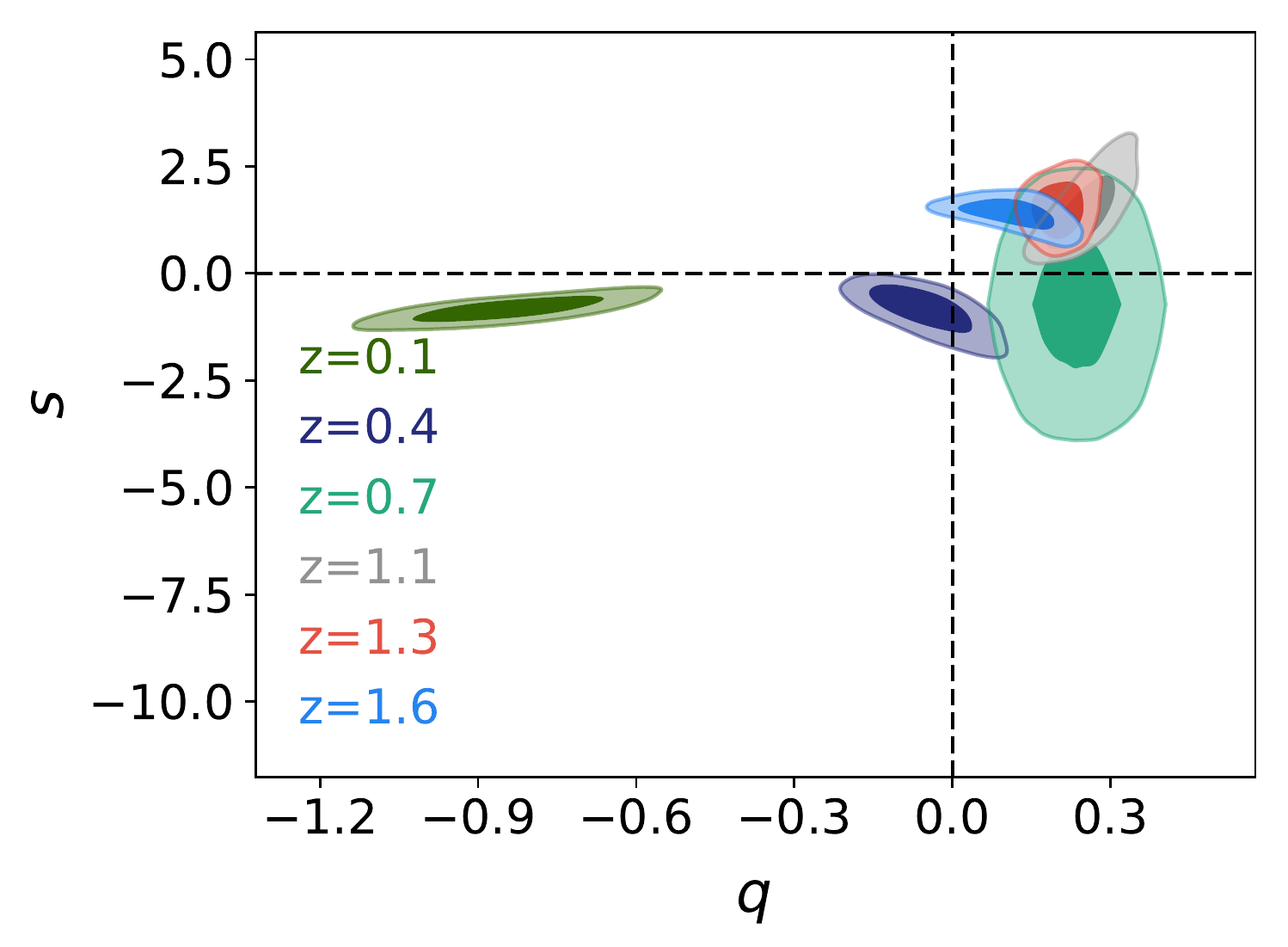}} 
\vspace{1mm} \resizebox{220pt}{180pt}{\includegraphics{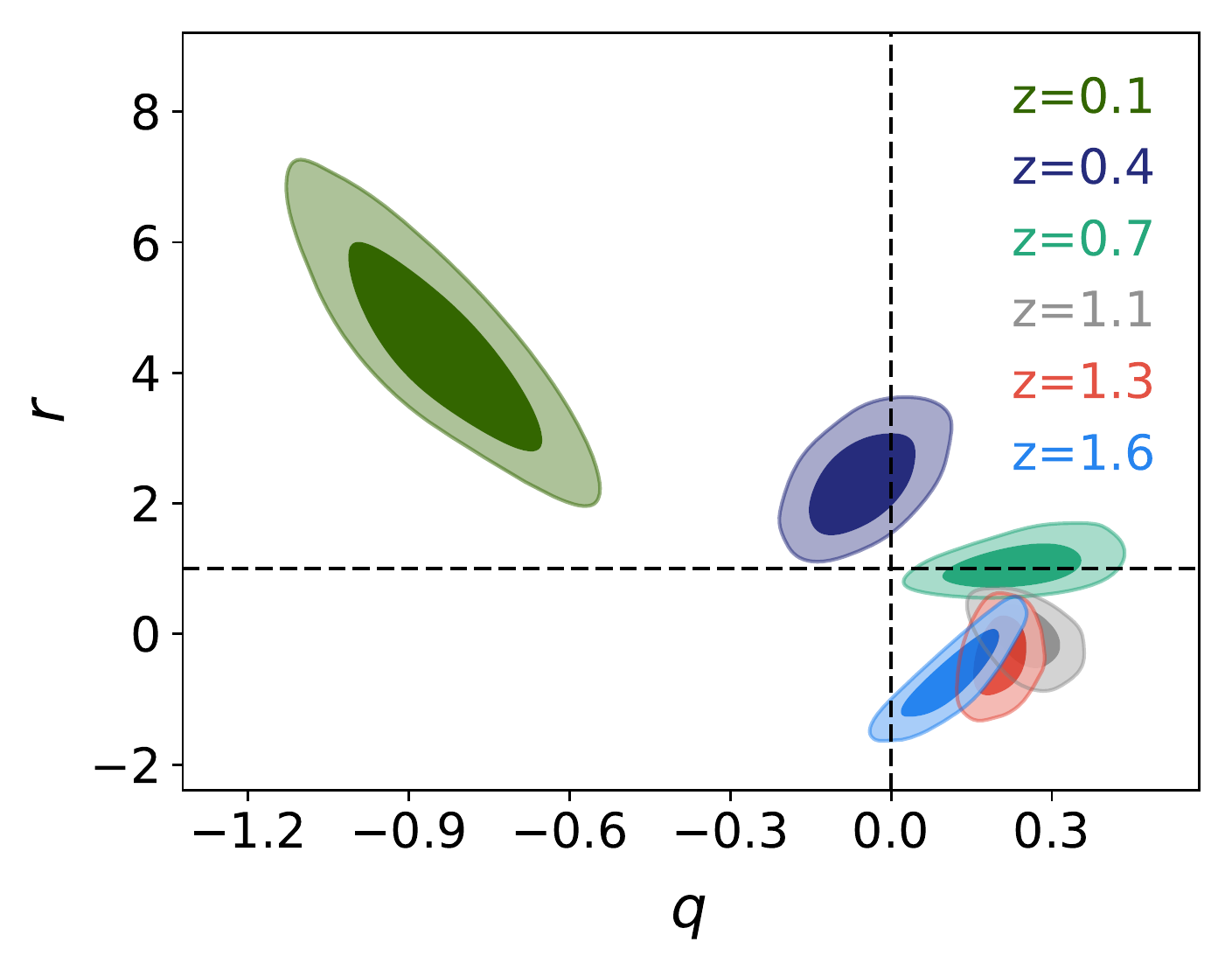}}
\end{center}
\caption{Different behaviours for the {\it statefinders} $(r,s)$ as well as its combination with deceleration parameter $q$.  Plot 1(Top, Left):  reconstructed behaviour of $r$ and $s$ as a function of redshift.  The innermost region is for $68\%$ confidence level whereas the outermost region is $95\%$ confidence level.The horizontal dashed line represents $r=1$ and horizontal solid line represents $s=0$. The two vertical lines represents the redshift range where $(r,s) = (1,0)$ is allowed.
Plot 2(Top, Right): evolution of allowed confidence contours in $(r,s)$ plane for different redshifts. Plot 3(Bottom Left): evolution of allowed confidence contours in $(s,q)$ plane for different redshifts. Plot 4(Bottom, Right): evolution of allowed confidence contours in $(r,q)$ plane for different redshifts. All the confidence contours are for $1\sigma$ and $2\sigma$ confidence levels and shading are same as in Figure 3.\label{fig:rs_z}}
\end{figure}

\begin{itemize}

\item The first observation from Figure 6 is that there is no $(r,s) = (1,0)$ fixed point for all redshifts ruling out the $\Lambda$CDM behaviour convincingly from low-redshift data. As one sees from the behaviour of  $(r,s)$ as function of redshift ( Top, Left in Figure 6), there is a redshift range between $0.5<z<1.0$ where $r=1$ and $s=0$ are allowed simultaneously and in this redshift range $\Lambda$CDM behaviour is possible. Outside this redshift range, $\Lambda$CDM behaviour is ruled out.

\item The confidence contours in $(r,s)$ plane for different redshifts (Top, Right in Figure 6) give important clue about allowed dark energy models. If one compares this figure with Figure (1a).\citep{state2},  one can easily conclude about possible dark energy behaviour allowed by the low-redshift data. It shows that in the past ( high redshifts), dark energy behaviours as given by different quintessence models ( including that with constant equation of state) are allowed, whereas around preset time ( low redshifts), models like Chaplygin Gas are suitable. In between, around $z=0.6-0.8$, when we expect the transition from deceleration to acceleration has taken place, $\Lambda$CDM is consistent with data.

The fact that no particular dark energy behaviour is consistent for the entire redshift range $0 \leq z \leq 2$ where low-redshift observations are present, is one of the most important results in this exercise and it poses serious challenge for dark energy model building that is consistent with low-redshift observations.

\item If one compares the contours in the $(r,q)$ plane (Bottom Right in Figure 6) with Figure 2 \citep{state2}, one can see that at low redshifts, the ``BRANE1" models ( as described  \citep{state2}) are consistent which results phantom type equation of state. At higher redshifts and in the decelerated regime ($q>0$), a class of brane-world models, called ``{\it disappearing dark energy}'' (refer to \citep{state2}) are possible that gives rise to transient acceleration. Hence, even if one models late time acceleration with various brane-world scenarios, it is not possible to fit the entire low-redshift range $(0 \leq z \leq 2)$ which we consider in our study, using a single brane-world description.

\end{itemize}

\subsection{ The Sound Speed}

In literature, there has been number of studies assuming dark energy to be a barotropic fluid where pressure of the dark energy is an explicit function of its energy density:

\begin{equation}
p_{DE} = f(\rho_{DE}).
\end{equation}

\noindent
Chaplygin and Generalized Chaplygin Gas, Van der Waals equation of state as well as dark energy with constant equation of state are few such examples which have been extensively studied in the literature. One interesting parameter associated with any barotropic fluid is its {\it sound speed}:

\begin{equation}
c_{s}^2 = \frac{dp}{d\rho}.
\end{equation}

To ensure stability of the fluctuations in the fluid, we should have $c_{s}^2 > 0$. Ensuring causality further demands $c_{s}^2 \leq 1$ (we refer readers to \citep{vik2} for situations where $c_{s}^2 > 1$ preserves causality). With a straightforward calculations, one can relate the {\it statefinder} $r$ with $c_{s}^2$ for barotropic fluid  \citep{state2, ss1, ss2}:

\begin{equation}
r = 1 + \frac{9}{2} (1+w_{eff}) c_{s}^2 = 1 + \frac{9}{2}\Omega_{DE}(1+w_{DE})c_{s DE}^2.
\end{equation}

\noindent
The first equality of the above equation relates the {\it statefinder} $r$ to the sound speed of total fluid of the Universe, $c_{s}^2 = \frac{dp_{T}}{d\rho_{T}}$ assuming that it is barotropic. This is relevant for unified models for Dark Sector where a single fluid describes both dark matter and dark energy \citep{ udm1, udm2, udm3, udm4, udm5}(see also  \citep{ vik3} for a UDM scenario with $c_{s}^2 = 0$ exactly). Here $w_{eff} = \frac{p_{T}}{\rho_{T}}$ as defined in section 3C (we ignore the contribution from baryons which is negligible compared to dark matter and dark energy and has negligible contribution to background evolution). 

\noindent
The second equality relates {\it statefinder} $r$ to the sound speed of the dark energy $c_{s DE}^2 = \frac{dp_{DE}}{d\rho_{DE}}$, assuming that the dark energy is described by a barotropic fluid.

Let us first discuss the possibility of an unified fluid for the dark sector  \citep{ udm1, udm2, udm3, udm4, udm5}. In Figure 6, we plot the redshift evolution of $r$ which shows that for $z<0.8$, $(r-1)>0$ and for $z>0.8$, $(r-1) < 0$. Moreover, the reconstructed $w_{eff}(z)$ as shown in Figure 4, shows that except around $z \sim 0$, where $w_{eff}$ can be phantom ($w_{eff}+1 < 0$),  $w_{eff} +1 >  0$ always. Putting these results in the first equality in equation (19), shows that for redshifts $z > 0.8$, $c_{s}^2 < 0$ and one can not have stable fluctuations in the unified fluid. This is unphysical as such unified fluid can not form structures in our Universe. Hence our results with low-redshifts observations show that the unified model for dark sectors with stable density fluctuations, is not compatible with the low-redshift observations. Behaviour of $c_{s}^{2}(z)$ for the total fluid is shown in Figure 7.

\begin{figure}
\begin{center}
\resizebox{260pt}{220pt}{\includegraphics{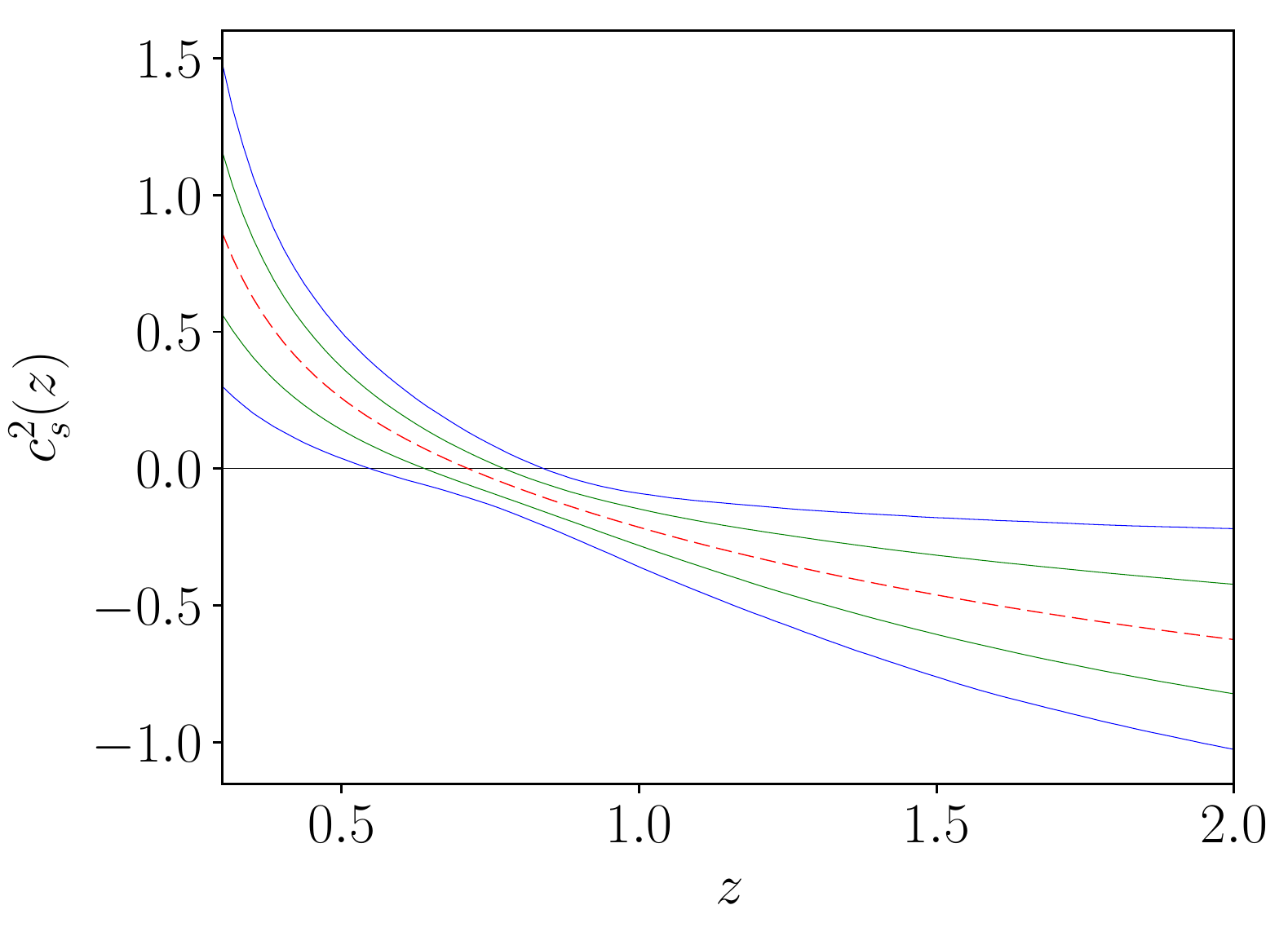}} 
\end{center}
\caption{Reconstructed $cs^2$ for the total fluid (ignoring the baryonic contribution) of the Universe as a function of redshift $z$ assuming the dark sector is governed by a single barotropic fluid. The innermost and outermost regions are same as in Figure 6.}\label{fig:cs_z}
\end{figure}

Next, we discuss the dark energy models as barotropic fluids and consider the second equality in equation (19). We already discuss the behaviour of $(r-1)$ for different redshifts in the previous paragraph. 
To reconstruct $c_{s DE}^2(z)$, we need to know the behaviour for $\Omega_{DE}$. But without that also, one can get an estimate of the $c_{s DE}^2 (z)$. For any physical dark energy model, $\Omega_{DE} >0$ always. Moreover, as shown in  Figure 5 for the reconstructed $w_{DE}$ for different $\Omega_{m0}$, $1+w_{DE} <0$ at low redshifts irrespective of the choice of $\Omega_{m0}$. Hence $c_{s DE}^2 <0$ at low redshifts for any dark energy barotropic fluid as constrained by low-redshift observations. We should stress that the dark energy dominates in the low-redshifts only and to have a consistent model for density perturbations in the Universe, we should take into account the perturbations in dark energy fluid even if it is negligible at some scales. But for barotropic dark energy fluid with $c_{s DE}^2 < 0$, perturbation in dark energy sector is unphysical and we can not have a consistent model for density perturbation. Hence the barotropic fluid model for dark energy as constrained by low-redshift observations, is also not consistent with stable solutions for growth of fluctuations.

%\begin{figure}[H]
%\begin{center} 
%\resizebox{220pt}{180pt}{\includegraphics{cs2_w.pdf}} 
%\hspace{1mm} \resizebox{220pt}{180pt}{\includegraphics{cs2_z.pdf}} 
%\end{center}
%\caption{Model independent constraints on sound speed and $w_{eff}$}\label{fig:cs2_plot}
%\end{figure}

\section{Conclusions}

Let us summarize the main results, we obtained in this study:

\begin{itemize}

\item We take a model independent cosmographic approach using $P(2,2)$ order Pade Approximation for the normalized Hubble parameter as a function of redshift and subsequently constrain the background evolution of our Universe around present time using low-redshift cosmological observations. We stress that our approach is independent of any underlying dark energy or modified gravity model, hence independent of amount of matter content in the Universe.

\item The constraint on jerk parameter at present time $j_{0}$, as well as the reconstructed {\it statefinders} $(r,s)$,  show that the $\Lambda$CDM behaviour is inconsistent with low-redshift data. 

\item Furthermore, the reconstructed total equation of state $w_{eff}$ of the Universe as constrained by low-redshift data is shown to be inconsistent with the same as constrained by Planck-2015 data for $\Lambda$CDM. This again confirms the apparent tension between low-redshift observations and the Planck-2015 results for $\Lambda$CDM.

\item With the combination of SNIa+BAO+TDSL+Masers data, we obtain the model independent constraint on $H_{0}$ that is fully consistent with the $H_{0}$ measurement by R16.

\item For the full combination of dataset (SNIa+BAO+TDSL+Masers+$H_{0}$+$H(z)$), our model independent measurement for $H_{0}$ is $72.56\pm 1.5$ Km/s/Mpc. This is in $3.46\sigma$ tension with Planck-2015 measurement for $H_{0}$ for $\Lambda$CDM.

\item Assuming that the late time acceleration is due to the presence of dark energy, we also reconstruct the dark energy equation of state $w_{DE}$ for three different choices of $\Omega_{m0}$. This reconstruction also shows the inconsistency with cosmological constant irrespective of the value of $\Omega_{m0}$.

\item The data allow only phantom model around present redshift. For $\Omega_{m0}=0.28$, multiple phantom crossing is evident. For higher values of $\Omega_{m0}$, pure phantom model is allowed although the overall shape of reconstructed $w_{DE}$ confirms the existence of phantom crossing. This rules out single field minimally coupled canonical scalar field models for dark energy by low-redshift observations. This result is purely model independent.

\item Reconstruction of the {\it statefinder diagnostics} $(r,s)$ shows that one single model for dark energy (e.g. quintessence, chaplygin gas or brane-world scenarios) can not explain the $(r,s)$ behaviours for the entire redshift range $0 \leq z \leq 2$ where low-redshift data are available. For different redshift ranges, different dark energy behaviours are allowed. This poses serious challenge to dark energy model building.

\item Finally using the constraints on the {\it statefinder} $r$, $w_{eff}$ and $w_{DE}$, one can get useful constraint on sound speed for the total fluid in the Universe as well as for the dark energy fluid assuming that they are barotropic. For both the cases, $c_{s}^2 < 0$ for certain redshift ranges. This gives rise to unstable perturbations in these fluids which is unphysical. So one can conclude that low-redshift data is not consistent with barotropic models for unified dark sectors  as well as with barotropic dark energy models.

\end{itemize}

To conclude, our model independent analysis with low-redshift data gives interesting insights for late time acceleration and, in particular,  for the underlying dark energy behavior. In particular, the available dark energy models may not be suitable to explain the low-redshift data and we may need more complex models  to explain the current set of data.
In other words, the approach seems promising because, being model independent, could really discriminate among the various proposals  in order  to remove the degeneration of $\Lambda$CDM where all models  converge at late epochs.  Finally the indication that a single fluid model, generally used  to account for the whole dynamics, seems  not sufficient: it could simply mean that  {\it coarse-grained} models are not realistic  in view to achieve a comprehensive  description of the Universe history.

Finally, in our study, we used the set of parameters $q_{0}, j_{0}, l_{0}, s_{0}$ (which are directly related to cosmography) instead of the actual parameters $P_{1}, P_{2}, Q_{1}, Q_{2}$ of the Pade approximation in Eq.(10). The results involving these two sets of parameters do not fully agree for redshifts $z >1.5$ as shown in Figure 2. So the results and tensions  for higher redshifts, ($z =1.5$ and above) quoted in our study, may change if one uses the set $P_{1}, P_{2}, Q_{1}, Q_{2}$ of parameters. But as shown in all the reconstructed plots for different cosmological quantities, the uncertainty in these quantities are already quite large for redshift $z=1.5$ and above. In conclusion,  one needs to be very careful about these facts while considering  results in the  redshift range $z=1.5$ and above.

\section{Acknowledgement}
SC acknowledges COST action CA15117 (CANTATA), supported by COST (European Cooperation in Science and Technology). SC is also  partially supported by the INFN sezione di Napoli (QGSKY). Ruchika  is funded by the Council of Scientific and Industrial Research (CSIR), Govt.~of India through Junior Research Fellowship. AAS acknowledges the financial support from CERN, Geneva, Switzerland where part of the work has been done. The authors thank Anto I. Lonappan for computational help. We acknowledge the use of publicly available MCMC code ``{\it emcee}" \citep{emcee}.


\begin{thebibliography}{199}
%(Ade et al. 2016)
\bibitem[\protect\citeauthoryear{Ade et al.}{2016a}]{ade1}
Ade~P.~A.~R. et al., 2016a, Astron.~Astrophys., {\bf 594}, A13 

\bibitem[\protect\citeauthoryear{Ade et al.}{2016b}]{ade2}
Ade~P.~A.~R. et al., 2016b, Astron.~Astrophys., {\bf 594}, A14 
 

%%(Alam et al. 2003)             
\bibitem[\protect\citeauthoryear{Alam et al.}{2003}]{state2}
Alam U., Sahni V., Saini T. D. and Starobinsky A. A., 2003, MNRAS, {\bf 344}, 1057 

%(Aviles et al. 2014 $\&$ Capozziello et al. 2018)(salvatore-2)
\bibitem[\protect\citeauthoryear{Aviles et al.}{2014}]{salv1}
Aviles A., Bravetti A., Capozziello S., Luongo O., 2014, Phys.~Rev.~D, {\bf 90}, 043531 


%(Babichev et al. 2008)
\bibitem[\protect\citeauthoryear{Babichev et al.}{2008}]{vik2}
Babichev E., Mukhanov V. and Vikman A., 2008, JHEP, {\bf 0802}, 101 



%(Babichev et al. 2008; Hu 2005 ; Sen 2006 ) (phcross-3)
\bibitem[\protect\citeauthoryear{Babichev et al.}{2008}]{phcross1} 
Babichev E., Mukhanov V. and Vikman A., 2008, JHEP, {\bf 02}, 101 


%(Bamba et al. 2009)
\bibitem[\protect\citeauthoryear{Bamba et al.}{2009}]{bamba}
Bamba K., Geng Chao-Qiang, Nojiri S., and Odintsov S. D.,  2009,  Phys. Rev. D {\bf 79}, 083014 

%(Barreiro $\&$ Sen et al. 2004; Burrage et al. 2011; Capozziello et al. 2005; Capozziello $\&$ Laurentis 2011; Dvali et al. 2000; Freese $\&$ Lewis 2002; Nicolis et al. 2009; Nojiri $\&$ Odintsov 2011 and  Nojiri et al. 2017 )(mod-9)

\bibitem[\protect\citeauthoryear{Barreiro et al.}{2004}]{mod1}
Barreiro T. and Sen A. A., 2004, Phys.~Rev.~D, {\bf 70}, 124013 

%(Benetti et al. 2017)
\bibitem[\protect\citeauthoryear{Benetti et al.}{2017}]{micol} 
Benetti M.,Graef L. L., Alcaniz J. S. ,2017,  JCAP {\bf 1704}, 003 

%(Bento et al. 2002; Billic et al. 2002; Davari et al.; Mishra $\&$ Sahni ;Scherrer 2004)(udm5)
\bibitem[\protect\citeauthoryear{Bento et al.}{2002}]{udm1}
Bento M. C., Bertolami O. and Sen A. A., 2002, Phys.~Rev.~D, {\bf 66}, 043507 

%(Betoule et al. 2014; Perlmutter et al. 1997; Riess et al.1998)    (sn-3)
\bibitem[\protect\citeauthoryear{Betoule et al.}{2014}]{sn1}
Betoule M. {\it et al.}, 2014,  [SDSS Collaboration],Astron.\ Astrophys.\  {\bf 568}, A22 

%(Beutler et al. 2011 , 2012; Blake et al. 2012; Lauren et al. 2013, 2014)  (bao-5)
\bibitem[\protect\citeauthoryear{Beutler et al.}{2011}]{bao1}
Beutler F. et al., 2011, Mon.Not.Roy.Astron.Soc., {\bf 416}, 3017 

\bibitem[\protect\citeauthoryear{Beutler et al.}{2012}]{bao2}
Beutler F. et al., 2012, Mon.Not.Roy.Astron.Soc., {\bf 423}, 3430


\bibitem[\protect\citeauthoryear{Billic et al.}{2002}]{udm2}
Billic N., Tupper G. B., Viollier R. D., 2002, Phys.~Lett.~B, {\bf 535}, 17 

\bibitem[\protect\citeauthoryear{Blake et al.}{2012}]{bao3}
Blake C. et al., 2012, Mon.Not.Roy.Astron.Soc., {\bf 425}, 405 


%(Bogdanos $\&$ Nesseris 2009; Nesseris $\&$ Shafieloo 2010) (ga-2)
\bibitem[\protect\citeauthoryear{Bogdanos $\&$ Nesseris}{2009}]{ga1}
Bogdanos C. and Nesseris S., 2009, JCAP, {\bf 0905}, 006 

%(Bonvin et al. 2017)
\bibitem[\protect\citeauthoryear{Bonvin et al.}{2017}]{holicow}
Bonvin V. et al., 2017, Mon.Not.Roy.Astron.Soc., {\bf 465}, 4914 


\bibitem[\protect\citeauthoryear{Burrage et al.}{2011}]{mod2}
Burrage C., De~Rham C., Heisenberg L., 2011, JCAP, {\bf 1105}, 025 

\bibitem[\protect\citeauthoryear{Davari et al.}{2018}]{udm3}
Davari Z. and Malekjani M. ,Artymowski M., 2018, arXiv:1805.11033 [gr-qc] 


  %``Improved cosmological constraints from a joint analysis of the SDSS-II and SNLS supernova samples,''
 

%(Capozziello et al. 2006)
\bibitem[\protect\citeauthoryear{Capozziello}{2006}]{emilio}
Capozziello S., Cardone V. F. , Elizalde E., Nojiri S. and Odintsov S. D., 2006,
  Phys.\ Rev.\ D {\bf 73},  043512 
  
\bibitem[\protect\citeauthoryear{Capozziello et al.}{2005}]{mod3}
Capozziello S., Cardone V. F. and Troisi A., 2005, Phys.~Rev.~D, {\bf 71}, 043503   

\bibitem[\protect\citeauthoryear{Capozziello et al.}{2018}]{salv2}
Capozziello S., D'Agostino R., Luongo O., 2018, JCAP, {\bf 1805}, 008 


\bibitem[\protect\citeauthoryear{Capozziello et al.}{2011}]{mod4}
Capozziello S., De Laurentis M., 2011,   Phys. Rept. {\bf 509}, 167 

%(Chiba $\&$ Nakamura 1998; Linder $\&$ Scherrer 2009)  (ss-2)
\bibitem[\protect\citeauthoryear{Chiba $\&$ Nakamura}{1998}]{ss1}
Chiba T.~and Nakamura T.,1998, Prog.~Theor.~Phys., {\bf 100}, 1077 
	

%(Deffayet et al. 2010)
\bibitem[\protect\citeauthoryear{Deffayet et al.}{2010}]{deff}
Deffayet C.~, Pujolas O.~, Sawicki I.~and Vikman A., 2010, JCAP, {\bf 1010}, 026 

\bibitem[\protect\citeauthoryear{Dvali et al.}{2000}]{mod5}
Dvali G. R., Gabadadze G. and Porrati M., 2000, Phys.~Lett., {\bf B 485}, 208 

%(Evslin et al. 2018)
\bibitem[\protect\citeauthoryear{Evslin et al.}{2018}]{evslin}
Evslin J.~,Sen A.~A, Ruchika,~2018, Phys.~Rev.~D, {\bf 97}, 103511 


%(Foreman-Mackey et al. 2013)
\bibitem[\protect\citeauthoryear{Foreman-Mackey et al.}{2013}]{emcee}
Foreman-Mackey D., Hogg D.~W.~, Lang D.~ and Goodman J.~, 2013, ``{\it Emcee: The MCMC Hammer}", Publ.~Astron.~Soc.~Pac., {\bf 125}, 306 

\bibitem[\protect\citeauthoryear{Freese et al.}{2002}]{mod6}
Freese K.and Lewis M., 2002, Phys.~Lett., {\bf B540}, 1 

%(Gao et al. 2016; Kuo et al. 2013; Reid et al. 2017)   (maser-3)
\bibitem[\protect\citeauthoryear{Gao et al.}{2016}]{maser1}
Gao F.~et al., 2016, Astrophys.~J., {\bf 817}, 128 

%(Gomez-Valent $\&$ Amendola 2018; Riess et al. 2018)(pantheon-2)
\bibitem[\protect\citeauthoryear{Gomez-Valent $\&$ Amendola}{2018}]{pantheon1}
Gomez-Valent A. and Amendola L., 2018, JCAP, {\bf 1804}, 051 


%(Gruber $\&$ Luongo 2014)
\bibitem[\protect\citeauthoryear{Gruber $\&$ Luongo}{2014}]{PA}
Gruber C. and Luongo O., 2014, Phys.~Rev.~D., {\bf 89}, 103506 

%(Heymans et al. 2013)
\bibitem[\protect\citeauthoryear{Heymans et al.}{2013}]{wl}
Heymans C.~et al., 2013, MNRAS, {\bf 432}, 2433 

\bibitem[\protect\citeauthoryear{Hu}{2005}]{phcross2}
Hu W., 2005, Phys.~Rev.~D, {\bf 71}, 047301 

%(Huterer $\&$ Starkman 2003)  
\bibitem[\protect\citeauthoryear{Huterer $\&$ Starkman}{2003}]{pc}
Huterer D.~and Starkman G.~, 2003, Phys.~Rev.~Lett., {\bf 90}, 031301 

\bibitem[\protect\citeauthoryear{Kuo et al.}{2013}]{maser2}
Kuo C.~et al., 2013, Astrophys.~J., {\bf 767}, 155 

\bibitem[\protect\citeauthoryear{Lauren et al.}{2013}]{bao4}
Lauren A. et al., 2013, Mon.Not.Roy.Astron.Soc., {\bf 427}, 3435 
\bibitem[\protect\citeauthoryear{Lauren et al.}{2014}]{bao5}
Lauren A. et al., 2014, Mon.Not.Roy.Astron.Soc., {\bf 441}, 24 

%(Lim et al. 2010)
\bibitem[\protect\citeauthoryear{Lim}{2010}]{vik3}
Lim E.~, Sawicki I.~ and Vikman A.~, 2010, JCAP, {\bf 1005}, 012 

\bibitem[\protect\citeauthoryear{Linder $\&$ Scherrer}{2009}]{ss2}
Linder E.~V.~and Scherrer R.~J.~, 2009, Phys.~Rev.~D, {\bf 80}, 023008 


%(Mehrabi $\&$ Basilakos)
\bibitem[\protect\citeauthoryear{Mehrabi $\&$ Basilakos}{2018}]{basilakos}
Mehrabi A. and Basilakos S.,  2018, arXiv:1804.10794 [astro-ph.CO] 

\bibitem[\protect\citeauthoryear{Mishra $\&$ Sahni}{2018}]{udm4}
Mishra S. S. and Sahni V., 2018, arXiv:1803.09767 [gr-qc]


\bibitem[\protect\citeauthoryear{Nesseris $\&$ Shafieloo}{2010}]{ga2}
Nesseris S. and Shafieloo A.,2010, MNRAS, {\bf 408}, 1879 

\bibitem[\protect\citeauthoryear{Nicolis et al.}{2009}]{mod7}
Nicolis A., Rattazzi R.and Trincherini E.,2009, Phys.~Rev.~D, {\bf 79}, 064036 


\bibitem[\protect\citeauthoryear{Nojiri et al.}{2011}]{mod8}
Nojiri S., Odintsov S. D.,2011,  Phys. Rept. {\bf 505}, 59  
\bibitem[\protect\citeauthoryear{Nojiri et al.}{2017}]{mod9}
Nojiri S., Odintsov S. D. and Oikonomou V. K., 2017, Phys.\ Rept.\  {\bf 692}1 


%(Padmanabhan 2003; Peebles and Ratra 2003; Sahni; Sahni and Starobinsky 2000)  (de -4)
\bibitem[\protect\citeauthoryear{Padmanabhan}{2003}]{de1}
Padmanabhan T., 2003, Phys. Rep. {\bf 380} 235 

%(Parkinson et al. 2012)
\bibitem[\protect\citeauthoryear{Parkinson et al.}{2012}]{lss}
Parkinson D. et al., 2012, Phys.~Rev.~D, {\bf 86}, 103518 

\bibitem[\protect\citeauthoryear{Peebles $ \&$ Ratra}{2003}]{de2}
Peebles P.~J.~E. and Ratra B., 2003, Rev. Mod. Phys. {\bf 75} 559 

\bibitem[\protect\citeauthoryear{Perlmutter et al.}{1997}]{sn2}
Perlmutter S. et al., Astrophys.~J., {\bf 483}, 565 (1997);

% (Pinho et al.) 
\bibitem[\protect\citeauthoryear{Pinho et al.}{2018}]{ohd} 
Pinho A. M. , Casas S. and Amendola L., 2018, arXiv:1805.00027[astro-ph.CO]

\bibitem[\protect\citeauthoryear{Reid et al.}{2017}]{maser3}
Reid M.~J.~et al.,2017, Astrphys.~J., {\bf 767}, 154 

% (Rezaei et al. 2017)  
\bibitem[\protect\citeauthoryear{Rezaei et al.}{2017}]{pade}
Rezaei M., Malekjani M., Basilakos S., Mehrabi A. and Mota D. F., 2017, Astrophys.~J., {\bf 843}, 65 

\bibitem[\protect\citeauthoryear{Riess et al.}{1998}]{sn3}
Riess A. G. et al., 1998, Astron.~J., {\bf 116}, 1009 

% (Riess et al. 2016)  
\bibitem[\protect\citeauthoryear{Riess et al.}{2016}]{R16}
Riess A. G. et al., 2016, Astrophys.~J., {\bf 826}, 56 

\bibitem[\protect\citeauthoryear{Riess et al.}{2018}]{pantheon2}
Riess A. G. et al., 2018, Astrophys.~J., {\bf 853}, 126 

%  (Riess et al.) 
\bibitem[\protect\citeauthoryear{Riess et al.}{2018}]{R18}
Riess A. G. et al., 2018, arXiv:1804.10655 [astro-ph.CO] 

\bibitem[\protect\citeauthoryear{Sahni}{2002}]{de3a}
Sahni V., 2002 astro-ph/0202076

\bibitem[\protect\citeauthoryear{Sahni}{2005}]{de3b}
Sahni V., 2005, astro-ph/0502032

% (Sahni et al. 2003)
\bibitem[\protect\citeauthoryear{Sahni et al.}{2003}]{state1}
Sahni V.,Saini T. D. ,Starobinsky A. A. and Alam U., 2003, JETP.~Lett., {\bf 77}, 201 

% (Sahni et al. 2008)
\bibitem[\protect\citeauthoryear{Sahni et al.}{2008}]{om}
Sahni V., Shafieloo A. and Starobinsky A. A., 2008, Phys. Rev. D, 78, 103502 

% (Sahni et al. 2014)
\bibitem[\protect\citeauthoryear{Sahni et al.}{2014}]{sahni}
Sahni V., Shafieloo A. and Starobinsky A. A., 2014, Astrophys.~J., {\bf 793}, L40 

\bibitem[\protect\citeauthoryear{Sahni et al.}{2000}]{de4}
Sahni V. and Starobinsky A.A., 2000, Int. J. Mod. Phys. {\bf D9} 373 


% (Scherrer)
\bibitem[\protect\citeauthoryear{Scherrer et al.}{2018}]{scherrer}
Scherrer R. J., 2018, arXiv:1804.09206 [astro-ph.CO] 


\bibitem[\protect\citeauthoryear{Scherrer et al.}{2004}]{udm5}
Scherrer R. J., 2004, Phys.~Rev.~Lett., {\bf 93}, 011301 


\bibitem[\protect\citeauthoryear{Seikel et al.}{2012}]{gp1}
Seikel M.~, Clarkson C.~and Smith M., 2012, JCAP, {\bf 1206}, 036 

\bibitem[\protect\citeauthoryear{Sen }{2006}]{phcross3}
Sen A. A., 2006, JCAP, {\bf 0603}, 010 

\bibitem[\protect\citeauthoryear{Shafieloo et al.}{2012}]{gp2}
Shafieloo A.~, Kim A.~G.~, Linder E.~V.~, 2012,Phys.~Rev.~D, {\bf 85}, 123530 


\bibitem[\protect\citeauthoryear{Valentino et al.}{2018}]{valentino}
Valentino E.~, Linder E.~ and Melchiorri A.~, 2018, Phys.~Rev.~D, {\bf 97}, 043528 

\bibitem[\protect\citeauthoryear{Vikman}{2005}]{vikman}
Vikman A., 2005, Phys.~Rev.~D, {\bf 71}, 023515 

\bibitem[\protect\citeauthoryear{Visser et al.}{2004}]{cosmography}
Visser M., 2004, Class.~Quant.~Grav., {\bf 21}, 2603 

\bibitem[\protect\citeauthoryear{Vitenti \& Penn-Lima}{2015}]{penn}
Vitenti S. D. P. and Penn-Lima M. ,2015, JCAP, {\bf 09}, 045 

\bibitem[\protect\citeauthoryear{Wei et al.}{2014}]{wei}
Wei H., Yan Xiao-Peng., Zhou Ya-Nan, 2014, JCAP, {\bf 1401}, 045 

\bibitem[\protect\citeauthoryear{Zhao et al.}{2017}]{gongbo}
Zhao Gong-bo et al., 2017, Nat.Astron., {\bf 1}, 627 

\end{thebibliography}
\end{document}